\documentstyle[11pt,aaspp4,tighten,flushrt]{article}

\eqsecnum
\def\apj{{ ApJ}}
\def\apjs{{ ApJS}}

\def\aa{{ AA}}
\def\aj{{AJ}}
\newcommand{\ea}{et al. \rm}
\newcommand{\mol}{\rm H_2}

\newcommand{\cp}{C$^+$}

\newcommand{\cc}{\rm cm^{-3}}
\newcommand{\icc}{\rm cm^{3}}

\received{29 January 1999}
\accepted{1 July 1999}
\lefthead{Kaufman et al.}
\righthead{Photodissociation Regions}

\begin{document}
\noindent\\To appear in {\it The Astrophysical Journal}\hfill\rm

\title{Far Infrared and Submillimeter Emission from Galactic and Extragalactic 
Photo-Dissociation Regions}

\author{Michael J. Kaufman}
\affil{Department of Physics, San Jose State University, San Jose, 
CA 95192-0106\\
and\\
Space Sciences Division, MS 245-3, NASA Ames Research Center, Moffett Field,
CA~94035} 
\author{Mark. G. Wolfire} \affil{Department of Astronomy, University
of Maryland, College Park, MD 20742} 
\author{David J. Hollenbach} \affil{Space Sciences Division, MS 245-3, NASA Ames Research Center, Moffett Field, CA~94035} 
\and 
\author{Michael L. Luhman\altaffilmark{1}}
\affil{Naval Research Laboratory, Remote Sensing Division, Code 7217, Washington, D.C. 20375}
\altaffiltext{1}{NRC-NRL Research Associate}

\begin{abstract}
Photodissociation Region (PDR) models are computed over a wide range of
physical conditions, from those appropriate to
giant molecular clouds illuminated by the interstellar radiation field to
the conditions experienced by circumstellar disks very close to hot massive 
stars. These
models use the most up-to-date values of atomic and molecular data,
the most current chemical rate coefficients, and the newest grain photoelectric
heating rates which include treatments of small grains and large molecules. In 
addition, we examine the effects of metallicity and cloud extinction on the
predicted line  intensities. Results are presented for PDR models with
densities over the range $n=10^1\,-\,10^7\,\rm cm^{-3}$ and for incident
far-ultraviolet radiation  fields over the range $G_0=10^{-0.5}\,-\,10^{6.5}$
(where $G_0$ is the FUV flux in units of the local interstellar value),
for metallicities $Z$=1 and 0.1 times the local Galactic value,
and for a range of PDR cloud sizes. We 
present line strength and/or line ratio plots for a variety of useful PDR 
diagnostics: [C~II] 158$\mu$m, [O~I] 63$\mu$m and 145$\mu$m, [C~I] 370$\mu$m 
and 
609$\mu$m, CO $J=1-0$, $J=2-1$, $J=3-2$, $J=6-5$ and $J=15-14$, as well as the 
strength of
the far-infrared continuum. These plots will be useful for the interpretation 
of Galactic and extragalactic far infrared and submillimeter spectra observable
with the {\it Infrared Space Observatory}, the {\it Stratospheric Observatory 
for Infrared Astronomy}, the {\it Submillimeter Wave Astronomy Satellite}, the 
{\it Far Infrared and Submillimeter Telescope} and other orbital and suborbital
platforms. As examples, we apply our results to ISO and ground based 
observations of M82, NGC 278, and the Large Magellenic Cloud. Our comparison of
the conditions in M82 and NGC 278 show that both the gas density and FUV flux
are enhanced in the starburst nucleus of M82 compared with the normal spiral
NGC 278. We model the high [C~II]/CO ratio observed in the 30 Doradus region 
of the LMC and find it can be explained either by lowering the average 
extinction through molecular clouds or by enhancing the density contrast 
between the atomic layers of PDR and the CO emitting cloud cores. The ratio
L[CO]/M[H$_2$] implied by the low extinction model gives cloud masses too
high for gravitational stability. We therefore rule out low extinction clouds
as an explanation for the high [C~II]/CO ratio and instead appeal to density
contrast in $A_V=10$ clouds. 
    
\end{abstract}


\section{Introduction}
In the past two decades infrared, submillimeter and millimeter observations 
of Galactic molecular clouds 
illuminated by ultraviolet radiation from nearby stars have revealed strong 
far-infrared (FIR) grain continuum emission as well as moderately strong 
emission from warm gas in the fine-structure lines of [C~II] 158$\mu$m, 
[O~I] 63$\mu$m and 145$\mu$m, [Si~II] 35$\mu$m, and [C~I] 370 and 609$\mu$m,
as well as  from low-$J$ transitions of CO.
A series of models have been used to interpret these 
observations (e.g., Tielens \& Hollenbach 1985a, hereafter TH85; van Dishoeck
\& Black 1986, 1988;  Sternberg \& Dalgarno 1989; Wolfire \ea 1990, 
hereafter WTH90; Hollenbach \ea 1991, hereafter HTT91; Abgrall \ea 1992; Le
Bourlot \ea 1993; K\"oster \ea 1994; Sternberg \& Dalgarno 1995; 
Draine \& Bertoldi 1996; Luhman \ea 1997a; Pak \ea 1998). 
Observations of Galactic sources are reviewed in Hollenbach \& Tielens (1997, 1999).
TH85 argued that the observed fine-structure line to dust-continuum flux 
ratios required a mechanism which was moderately efficient ($\sim 0.1-1\%$)
at converting far-ultraviolet 
(FUV) flux into atomic and molecular gaseous line emission.  
They showed that this condition is satisfied by grain photoelectric heating 
in the regions  where FUV radiation from hot stars impinged on 
opaque molecular clouds. TH85 dubbed regions where the FUV radiation played
a significant role in the heating and/or the chemistry ``photodissociation
regions'' (PDRs), and they showed that the emission from such regions could be
primarily parameterized by the cloud density, $n$, and the 
strength of the FUV ($6$eV$\,<h\nu\,<\,13.6$eV) radiation field, $G_0,$
 illuminating the cloud (where $G_0$ is in units of the ``Habing 
Field'', $1.6\times 10^{-3}\,\rm erg\,cm^{-2}\,s^{-1}$). In particular, 
WTH90 showed that far-infrared observations of PDRs would be extremely useful
in determining the physical properties of these regions.  

PDRs are the origin of much of the infrared radiation from the interstellar
medium (ISM). The incident starlight is absorbed by dust grains and
large carbon molecules (polycyclic aromatic hydrocarbons, or PAHs) and 
reradiated primarily as PAH infrared features and infrared continuum radiation.
As much as 
0.1-1\% of the absorbed starlight, however, is converted to gas heating via
photoelectric ejection of electrons from grains. The gas 
generally attains a higher temperature than the grains because of the much less
efficient cooling of the gas (often via [C~II] 158$\mu$m and [O~I] 63$\mu$m
emission) relative to the radiative dust continuum cooling, and inefficient 
coupling between gas and grains means that the 
gas and grains remain at different temperatures.
Much of the [C~II] and [Si~II] and most of the [O~I], [C~I] and CO rotational 
emission from the Milky Way Galaxy arises from PDRs.

PDRs include all the neutral gas in the ISM where FUV photons dominate a 
significant aspect of the chemistry and/or heating. Traditionally, PDRs have 
been associated with atomic gas. However, with the above definition, PDRs also 
include material in which the hydrogen is molecular and the carbon is mostly in
CO, but where FUV flux still strongly affects the chemistry of oxygen and 
carbon not locked in CO (photodissociating OH, O$_2$ and H$_2$O, for example)
and affects the ionization fraction as well. The transition from H to H$_2$ 
and from 
C$^+$ to CO occurs in PDRs. With the exception of the molecular gas in dense,
star-forming cores, most molecular gas in the Galaxy is found at A$_V\lesssim
5$ in giant molecular clouds where FUV radiation still plays an important role
(Wolfire, Hollenbach \& Tielens, 1993). 
Therefore, neutral diffuse clouds ($A_V<1$), translucent clouds ($A_V\sim 1$),
as well as 90\% of molecular clouds ($A_V>1$) are PDR gas - thus, most of the 
ISM is in PDRs. 

In addition, the warm neutral medium ($n\simeq 0.25\,\rm cm^{-3}$) of the 
Galaxy (c.f. McKee \& Ostriker 1977; Wolfire \ea 1995), the neutral gas around
planetary nebulae,
and photodissociated winds from red giant or AGB stars are PDRs. Densities in 
PDRs range from $n=0.25\,\rm cm^{-3}$ to $n\gtrsim 10^7\,\rm cm^{-3}$, and 
incident FUV fluxes range from $G_0\lesssim 1$ (the interstellar
radiation field) to $G_0\gtrsim 10^6$ (the field $\lesssim 0.1$pc from an O
star). We treat the range $10\,{\rm cm^{-3}}\,\le\,n\,\le\,10^7\,\rm cm^{-3}$ 
and $0.3\,\le\,G_0\,\le\,3\times 10^6$ in this paper. Comparison of 
observations
with our models not only provide a diagnostic of $n$, $T$ and $G_0$ from the
emitting regions, but also constrain the metallicity, the geometry, the dust
properties and (for regions smaller than the beam) the cloud or clump 
properties
such as the size, volume and area filling factors, and number of clouds or
clumps in the beam. 
 
As in Galactic sources, observations of PDRs in external galaxies show
that the fine structure lines of [C~II] and [O~I] generally dominate
the global gas cooling (e.g., Stacey et al. 1991; Madden et al. 
1993; Carral \ea 1994; Poglitsch et al.  1995; Lord et al. 1997), but 
[C~I] and CO emission
lines from well-shielded PDR gas can also contribute significantly (e.g.,
B\"{u}ttgenbach et al. 1992; White et al. 1994; Stutzki et al. 1997). 
The measured strengths of especially the [C~II] and [O~I] lines and the FIR
continuum, when
compared to PDR models, can provide effective diagnostics of
the typical conditions in extragalactic
interstellar media (e.g., WTH90, Carral et
al. 1994; Lord et al.  1997). In galaxies, however, the derived physical
parameters are uncertain to the extent that media other than opaque PDRs can
contribute to the line emission.  For example, shock-heated media may
contribute significantly to the [OI] 63 $\mu$m emission (Draine, Roberge,
\& Dalgarno 1983;  Hollenbach \& McKee 1989), as observed in the starburst
galaxy NGC 6240 (Lord et al. 1997; Luhman et al. 1997b).  Likewise, the
extended low-density cold neutral component
of the ISM (which is a non-opaque or diffuse PDR) 
may dominate the large-scale [C~II] 158 $\mu$m emission, as 
suggested by Madden \ea (1993) for the galaxy NGC 6946. Furthermore, diffuse
HII regions may dominate the production of [Si~II] or [C~II] in some galaxies
(Lord \ea 1994; Lord \ea 1996; Heiles 1997).

New and upcoming infrared and submillimeter airborne and space-based platforms
and sensitive detectors/receivers are greatly enhancing the sensitivity of
extragalactic PDR observations.  As a result, the number and variety of
observed extragalactic PDRs continue to grow, posing increasingly more
complicated challenges to theoretical PDR models.  For
instance, recent [C~II] and [O~I] FIR spectra of normal (Lord et al. 1996;
Malhotra \ea 1997) 
and starburst (Fischer et al. 1996; Satyapal et al. 1999) galaxies
obtained with the Infrared Space Observatory (ISO) reveal PDR gas that is
generally typical of that found in other Galactic and extragalactic PDRs,
where the ratio of [C~II] to FIR flux is $\sim$
$(1-3)\times 10^{-3}$ (Stacey et al. 1991). However, for some sources such as
ultraluminous infrared galaxies (Luhman et al. 1998; Fischer et al. 1999) 
and a few normal galaxies (Malhotra et al. 1997), ISO data indicate
extremely low [C~II]/FIR ratios $\sim (1-5)\times 10^{-4}$.  Such low ratios
may  indicate that the sources of the
grain heating are not OB stars, but are objects which radiate considerable 
bolometric luminosity in bands outside the FUV; or these ratios may indicate 
unusually high ratios of $G_0/n$, which result in a dramatic decrease in the 
grain photoelectric heating of the gas because of the increase of positive 
grain charge (c.f. \S 3). Conversely, in dwarf irregular galaxies, [C~II]/FIR
is much higher than in the Galaxy (e.g., [C~II]/FIR$\sim$1\% in the Large
Magellanic Cloud; Poglitsch et al. 1995; Israel et al. 1996; Pak \ea 1998) 
perhaps due 
to the lower metallicities and lower dust-to-gas ratios. The paucity of dust that
characterizes low metallicity galaxies increases the penetration depth of
photodissociating UV photons and produces clouds with reduced-sized CO cores
surrounded by larger PDRs, as evidenced by high ratios of [C~II]
to CO (Poglitsch et al. 1995; Israel et al. 1996;  Madden et al.  1997). 
Likewise, recent sensitive AST/RO observations of [C~I] in the LMC show
that the extent of [C~I] emission, created by the photodissociation of CO, is also
enhanced (Stark et al. 1997).  We note that such metal-deficient galaxies are 
not good
candidates for the standard conversion of CO intensity to H$_2$ mass (Maloney 
\& Black 1988; Sakamoto 1996).  Several authors have constructed PDR models
appropriate to the Magellanic Clouds, with particular emphasis on the 
C$^+$/C/CO transition (Maloney \& Black 1988; van Dishoeck \& Black 1988;
Lequeux et al.  1994;  Maloney \& Wolfire 1997; Pak \ea 1998).  These studies
demonstrate that in order to match the observed PDR line emission from
metal-deficient systems such as the Magellanic Clouds, low-abundance
models specific to the environment are required.

WTH90 presented especially useful contour plots of the fine-structure emission
and line ratios predicted by their PDR models as a function of $G_0$ and $n$. 
Several recent developments motivate us to update the calculations presented 
by WTH90. From a theoretical standpoint, many of the important physical 
parameters which effect the calculation of emissions from PDRs (e.g. collision 
strengths, chemical rate coefficients, photoelectric heating rates) have been 
revised in recent years, in some cases by orders of magnitude, so that we might
expect the  predictions of these models to change considerably. 
We extend the work of WTH90 to include a larger parameter space for $n$ 
and $G_0$, and to include a study of the effects of changing the metallicity 
and geometry.
We also include results for a wider variety of predicted diagnostic lines
(e.g. [C~I] and CO). From an 
observational standpoint, results from recent, existing, and future observatories will
be used to study Galactic and extragalactic PDRs in greater detail than ever 
before. These include the {\it Infrared Space Observatory} (ISO);
the {\it Stratospheric Observatory for Infrared Astronomy} (SOFIA) [NASA's
planned successor to the {\it Kuiper Airborne Observatory} (KAO)], which will 
be used to  
study far-infrared sources with unprecedented spectral and spatial resolution;
the {\it Submillimeter Wave Astronomy Satellite} (SWAS) and the Odin telescope,
which will be used to 
observe the sky in submillimeter transitions of CI, O$_2$, H$_2$O and CO; and 
the {\it Far Infrared and Submillimeter Telescope} (FIRST), which will be used 
to study a wide range of Galactic and extragalactic phenomena over a broad 
spectral range. 

The outline of this paper is as follows:
In \S 2 we discuss the relevant details of our PDR model. 
We concentrate on the important differences between our present
calculations and previous studies by our group. Recent observational work
has led us to select a new set of atomic abundances, while recent theoretical
work has led to improvements in our understanding of several important 
reaction and heating rates, and we discuss these and other important 
differences.

In \S 3 we present a series of contour plots of line strengths and line 
ratios of important atomic and molecular cooling species.
We divide an opaque PDR into a warmer surface region ($A_V\lesssim2$) and
a cooler, more molecular interior ($A_V\sim 1-5$). 
In \S 3.1 and \S 3.2 we focus on emission from the surface regions and present 
surface temperatures as well as line strengths and line ratios
of several important far-infrared cooling lines including 
the [C~II] 158$\mu$m line, and the [O~I] 63$\mu$m and 145$\mu$m lines. 
We also compare 
the strengths of these transitions with the integrated far-infrared continuum
intensity expected from PDRs. This line-to-continuum ratio provides a 
measure of the efficiency with which stellar photons are converted to gas 
heating in PDRs. In \S 3.3 we concentrate on the interior part of the PDR
where longer wavelength tracers originate: the 
neutral carbon transitions at 609$\mu$m and 370$\mu$m, as well as several 
millimeter and submillimeter transitions of carbon monoxide.

In \S 4, we discuss how changes in the assumed metallicity and in the model 
geometry can affect our results. We briefly describe our procedure (WTH90)
for deriving the average physical parameters of an ensemble of PDRs when 
large beams are used to study distant (e.g. extragalactic) interstellar media.
Our procedure for applying our results to
ensembles of clouds is relevant for the interpretation of the far infrared,
millimeter and submillimeter observations of, for instance, the nuclei of 
starburst galaxies which exhibit metallicities close to solar, as well as 
the Magellenic Clouds or other irregular galaxies which show reduced 
metallicities.
These results are applied to the starburst galaxy M82, the normal spiral
galaxy NGC 278, and to the Large Magellenic Cloud.   

\section{PDR Models}

The basis for our new modeling effort is the PDR model of TH85 and HTT91,
which computes
a simultaneous solution for the chemistry, radiative transfer, and thermal 
balance in PDRs. Assuming gas phase elemental abundances and grain properties,
each model is essentially described by a constant 
density of H nuclei, $n$, throughout the PDR and the intensity of the 
far-ultraviolet radiation field at the PDR surface, $G_0$. 

These models have been used in the past to 
show how the structure of a molecular cloud varies with
increasing optical depth. TH85 and HTT91 described the general morphology 
of the surface
layers for typical ratios $G_0/n\gtrsim10^{-2}\,\rm cm^3$, as is often the
case:  an outer layer ($A_V\sim 1$) of atomic H, C$^+$ and O, a central
layer where the H makes the transition to $\rm H_2$ ($A_V\sim 1-2$) and 
where C$^+$ becomes
C and then CO ($A_V\sim 2-4$), and an inner layer ($A_V\sim 3-10$) where 
the oxygen not incorporated into CO is still mainly in atomic O. At
larger depths, we leave the PDR and enter the very opaque molecular cores 
where the chemistry and heating are dominated by cosmic rays. 

\subsection{The Standard Model}
We list here the standard input parameters for the model results presented 
in the contour plots of the following section. The input parameters are: the 
micro-turbulent velocity dispersion of the gas, $\delta v_D$; the gas-phase 
elemental abundances;
the dust abundances, including those of polycyclic aromatic hydrocarbons (PAHs)
which generally dominate the grain-photoelectric heating; the gas 
density, $n$; and the strength of the incident FUV field, $G_0$.

Our model clouds are assumed to be pressure supported with a 
microturbulent Doppler line width of $\delta v_D=1.5$ $\rm km\,s^{-1}$; 
this value of the line width is taken from observational studies of 
fine-structure and
CO line widths (e.g. Hollenbach \& Tielens 1999) and is appropriate for the 
majority of cases which
we consider. This value of $\delta v_D$ is low for cases of high $G_0$ and 
high for cases of low 
$G_0$, but variation of this parameter by factors of $\lesssim$2 has little 
effect on the calculated line strengths (Wolfire, Hollenbach \& Tielens 1989).
We use the cool diffuse cloud gas phase abundances from Savage \& Sembach 
(1996) for our standard model.
In \S 4, we discuss the effects of
changes in these gas-phase abundances. We assume the same grain FUV absorption
characteristics as used in TH85 and HTT91, but we modify the grain 
photoelectric heating rates and the formation rate of $\rm H_2$ on grains, as 
described below. With these parameters fixed, our study
covers the range $10\,{\cc}\,\le\,n\,\le\,10^7\,\cc$ and 
$10^{-0.5}\le\,G_0\,\le\,10^{6.5}$. Table 1 summarizes the standard input 
parameters. 

Besides changing the standard gas phase abundances (note in 
particular the decrease in the gas phase carbon abundance in Table 1 from the 
value used in WTH90), we have made a number of 
changes to the chemical, heating and cooling rate coefficients as compared with
TH85, WTH90 and HTT91.
%Wolfire Contribution%
We use revised collision rates as discussed in Burton et al.\ (1990), 
Spaans et al.\ (1994), and 
Wolfire et al.\ (1995).  These include rates for the collisional 
excitation of H$_2$ by impacts with  
H, H$_2$, and $e^-$ (using modified rates of 
Draine, Roberge, \& Dalgarno 1983 and Hollenbach \& McKee 1979 as discussed in 
Burton et al.\ 1990), the excitation of O I by impacts 
with H$^0$, ${\rm e^-}$,  and H$^+$ (P\'equignot 1990), 
the impacts of
C II with  ${\rm e^-}$(Blum \& Pradhan 1992), and  C and O collision rates with
H$_2$ (Schr\"oder et al.\ 1991; Jaquet et al.\ 1992). The ortho and
para H$_2$ fractions needed for the collision rates are calculated as in 
Burton et al.\ (1992). 

We use the grain photoelectric heating rate
as derived by Bakes \& Tielens (1994). This rate includes a size
distribution of particles extending from large grains ($\sim 0.25$ $\mu$m)
to small ($\sim 5$ \AA ) PAHs and explicitly accounts for the 
microphysics of small particles.  We use the ``Flat'' PAH geometry
as discussed in their paper, and a grain size distribution proportional to 
$a^{-3.5}$ (the so-called MRN distribution [Mathis, Rumpl \& Nordsieck 1977]) 
which extends to the PAH domain.  The total PAH abundance 
$n_{\rm PAH}/n$ is approximately $4.0\times 10^{-7}$ with this assumption. 
Also included is the Bakes \& Tielens
(1994) cooling term due to electron recombinations onto positively charged 
grains. Vibrational heating by FUV pumped H$_2$ as well as the vibrationally
excited H$_2$ (H$_2^*$) 
abundances and H$_2$ rotational level populations 
are calculated using the prescription given in Burton et al.\ (1990).
The dust temperature, important for gas heating at large optical depths,
is found using the formulae in HTT91,  and
the gas-grain heat exchange is calculated as in Hollenbach \& McKee (1989).

The gas phase chemical reaction rates have been updated 
according to the list of Millar, Farquhar, \& Willacy (1997; UMIST Database).
We have incorporated PAH chemistry using 
PAH$^-$, PAH$^0$, and PAH$^+$ species and have 
included photoionization and electron recombination onto PAHs,
charge exchange between metal ions and PAH$^0$, and metal ion recombination 
with PAH$^-$. The chemical rates are calculated using the
equations in Bakes \& Tielens (1994) and Draine \& Sutin (1987). 
We have also modified the cosmic ray heating and ionization terms using fits 
to the results of Shull \& Van Steenberg (1985) (see Wolfire et al. 1995 for
functions). 
We have retained most of the photodissociation rates as in the Wolfire
et al.\ (1990) paper with a few notable exceptions. The rate for OH
photodissociation is taken from the Millar et al.\ (1997) list.
The photodissociation rate of O$_2$ and the photoionization rate of Mg
are taken from Spaans et al.\ (1994)
for the $T_{\rm eff}=30,000$ K case, appropriate for the interstellar FUV
radiation field. Abgrall et al.\ (1992) reevaluated the dissociation rate of 
H$_2$ in the interstellar medium. Their rate for an unshielded H$_2$ molecule
in the radiation field of the local ISM amounts to $R_{\rm photo}=4.7\times 
10^{-11}$~s$^{-1}$ and we have adopted this rate. The photodissociation of CO 
is calculated as in  Burton et al.\ (1990) except for the CO shielding 
factor which is  taken from van Dishoeck \& Black (1988). This factor 
includes both H$_2$ shielding and CO self-shielding. 

Previously, we and others have used relatively complex expressions (HM79, HM89)
for the rate coefficient of $\mol$ formation on grain surfaces. This rate 
coefficient depends sensitively on the H
atom collision rate with grains, the sticking coefficient (itself a 
function of the grain temperature and the thermal velocity of the H atoms),
and the probability of an absorbed atom producing an $\mol$ molecule before
evaporating. While the collision frequency increases with the gas temperature,
the sticking coefficient and $\mol$ formation efficiency decrease with 
temperature.  
In light of the realistic uncertainties in the H$_2$ formation rate, we have 
simplified our former treatment.  Here we adopt the observationally derived
(c.f. Jura 1974) formation rate coefficient of 
$R_{form}=3\times 10^{-17}$ cm$^3$ s$^{-1}$. The formation rate per unit
volume is then given by $R_{form}n_{\rm HI}n$. We find that the resultant
formation rate coupled with the photodissociation rate of Abgrall et al.\
(1992) closely matches the observed molecular fraction as a function of total
hydrogen column density as seen in the diffuse interstellar medium (e.g.,
Dixon, Hurwitz, \& Bowyer 1998). 

We find that the most significant differences in the [C~II] and [O~I]
line intensitites from those reported in the Wolfire et al. (1990) paper 
result from the new heating term which now includes the important effects of 
PAH heating. At high densities and FUV fields, the modified OI collision rates
are also partly responsible for the differences. We also find that PAH 
chemistry can affect the depth of the \cp/C/CO transition with PAHs drawing the
transition closer to the PDR surface (see Bakes \& Tielens 1998 for a 
discussion of the effects of PAHs on the chemistry of PDRs). Neglecting PAH
chemistry can lead to underestimates in the [C~I] line intensities by a factor
2.

\section{Results}
In this section, we show our results for the strengths and line ratios
of various cooling lines, as well as the continuum emission from grains, 
over
the complete parameter space {\it for our standard model} (c.f. Table 1). 
We present contour plots over the range
$10\,{\rm cm^{-3}}\le\, n\,\le\,10^7\,{\rm cm^{-3}}$ and 
$10^{-0.5}\le\,G_0\,\le\,10^{6.5}$. 
We begin by presenting contours of the PDR surface temperature as a function of
$G_0$ and $n$ (\S 3.1). In \S 3.2,
we give results for the strengths of the important far infrared lines 
[C~II] 158$\mu$m, [O~I] 63$\mu$m and 145$\mu$m, and the far-infrared 
continuum; in \S 3.3, we present the strengths of the millimeter and 
submillimeter transitions [C~I] 370$\mu$m and 609$\mu$m, CO $J=1-0$, $J=2-1$,
$J=3-2$, $J=6-5$, and the far-infrared CO $J=15-14$ transition. Table 2 lists
these ten important PDR cooling transitions, their wavelengths, upper energy
levels $E_{upper}$, and  critical densities $n_{cr}$. 

\subsection{Thermal Balance and the PDR Temperature}
Unlike Wolfire et al. (1990), who found in many cases a significant initial 
rise in gas temperature with cloud depth, we find that the inclusion of PAH 
heating leads to a relatively constant PDR gas temperature from the cloud 
surface to a depth where either the heating or cooling changes significantly 
and the temperature drops. The heating is generally 
dominated by the grain photoelectric effect, and this changes primarily when 
dust attenuation significantly lowers the FUV flux, at $\tau_{FUV}>1$ or a 
column density $N\gtrsim
10^{21}\,\rm cm^{-2}$. Collisional deexcitation of FUV-pumped $\rm H_2$ can 
dominate the heating in cases of high $G_0\gtrsim 10^3$ {\it and} high $n\sim
10^5-10^6\,\rm cm^{-3}$. The cooling changes when the primarily atomic (H, C$^+$
and O) surface becomes molecular (H$_2$ or CO) at depths of $N\gtrsim
3\times 10^{21}\,\rm 
cm^{-2}$ for $G_0/n\gtrsim\,3\times 10^{-3}\,\rm cm^3$ (such that dust 
shields the molecules from dissociating radiation) or at smaller column 
densities for $G_0/n\lesssim \,3\times 10^{-3}\,\rm cm^3$ (such that 
self-shielding by H$_2$ and CO becomes dominant).

Figure 1 shows the PDR surface temperature, $T_S$, as a function of $G_0$ and
$n$. In general, $T_S$ rises with $G_0$ as would be expected since grain 
photoelectric heating dominates and the FUV flux is the source of the heating.
There is an exception to this rule for very high fluxes, $G_0\sim\,10^6$, and
very low gas densities [$n\lesssim\, 10^2\,\rm cm^{-3}$], where the grains 
become so positively charged due to the FUV flux that increasing $G_0$ actually
decreases the net heating. This heating drop stems from the poor heating 
efficiency of ejecting electrons from positively charged dust grains and 
increased rate of recombination cooling with the grains. However, as we 
discuss below, 
the results in this corner of parameter space are not realistic since the model
does not take into account grain drift through the gas due to radiation 
pressure, which becomes important in this region. Fortunately, there are not 
many astrophysical regions which correspond to the upper-left quadrant of our 
$G_0-n$ parameter space. 

The behavior of $T_S$ with density is somewhat more complicated. At low 
densities ($n\lesssim\, 10^4-10^5\,\rm cm^{-3}$), the temperature drops for
increasing $n$ at fixed $G_0$. This behavior arises because the cooling is 
generally dominated by the [C~II] 158 $\mu$m and [O~I] 63 $\mu$m transitions
whose critical densities lie at 
$n_{cr}({\rm C~II})\simeq 3\times 10^3\,\rm cm^{-3}$ and 
$n_{cr}({\rm O~I})\simeq 5\times 10^5\,\rm 
cm^{-3}$, respectively. Thus the cooling rate per unit volume increases with 
$n^2$ in the low density regime, whereas the grain photoelectric heating 
increases with a lower density dependence. The heating has, at least, an 
$n^1$ dependence since the grain density is proportional to $n$. In addition, 
as the gas phase electron density increases, the grains become 
less positively charged. The diminished
charge leads to a larger fraction of photoelectrons which escape the grains 
and increases the kinetic energy which they supply to the gas. The total 
density dependence for grain heating, however, is less than the $n^2$ 
dependence of cooling. 
At moderate densities ($n\sim\,\rm 10^4-10^6\,\rm cm^{-3}$), [O~I] and
[C~II] attain LTE, the cooling rate per unit volume is proportional to $n$, and
the heating rate increases faster than the cooling rate. Here, $T_S$ increases
with increasing $n$ at fixed $G_0$. Because the 
interstellar cooling curve is relatively flat as a function of $T$ for 
$300\,{\rm K}<T<3000\,\rm K$ 
(Dalgarno \& McCray 1972), the temperature rapidly rises with $n$ in this 
regime ($G_0\gtrsim 10^4$) to thousands of degrees, where coolants 
such as [O~I] 6300\AA~can
become important and balance the heating. Finally, at very high densities
($n\gtrsim \,10^6\,\rm cm^{-3}$), and especially for $G_0\lesssim\,10^4$, the
temperature again declines with $n$. Here the grains have become neutral so 
that the heating rate per unit volume due to grains is just proportional to 
$n$ (or the grain density), whereas the cooling increases somewhat more 
rapidly with $n$ because of the presence of additional coolants such as [O~I]
6300\AA~ and gas collisions with cold dust grains, whose efficiencies
go as $n^2$,  
in addition to the LTE coolants [O~I] 63$\mu$m and [C~II] 158$\mu$m whose 
efficiencies go as $n$.
   
\subsection{[C~II] 158$\mu$m, [O~I] 63$\mu$m, 145$\mu$m and the Far-Infrared
Continuum}
In this section we present results for the strengths of
several important far-infrared tracers of photodissociation regions. The 
emission is produced primarily in the outer regions of the UV illuminated 
clouds, where $\mol$ and CO are photodissociated and cooling is dominated 
by neutral species and by singly ionized species whose ionization
potentials are less than 13.6 eV. It is this outer region 
where the temperature contour plot (Fig.\ 1) applies
and where the PDR line cooling occurs through the [C~II] and [O~I] 
transitions.

In Figure 2, we show schematically the various regimes which determine the 
relative strengths of the principal coolants of PDRs: the [C~II] 158$\mu$m 
transition and the [O~I] 63$\mu$m transition. Above the line marked 
``$v_{drift}=v_{turbulent}$''
radiation pressure drives grains through the gas at velocities 
greater than the assumed average turbulent velocities ($\sim \delta v_D$) in
the gas. The models are therefore
inconsistent far above this line since the models assume no significant
drifting of grains with respect to the gas, and no significant gas heating or
chemical  effects due to the grain drift. The line marked $n=n_{cr}^{\rm C}$
is roughly the critical density of [C~II] 158$\mu$m, CO $J=1-0$ and 
[C~I] 370$\mu$m. To the 
left of this line, the line intensities of these species rise rapidly 
with $n$ (linearly with $n$ for optically thin transitions at constant column
density and temperature). To the right 
of this line, the line 
intensities increase more slowly with $n$ (as $n^0$ for constant column 
density and temperature as
LTE is reached). The line marked $T_S=92\,\rm K$ marks the regions with
surface temperatures of 92~K, the excitation temperature above ground 
of the [C~II] 158$\mu$m transition. Above this line, the PDR temperatures 
exceed 92~K and the [C~II] 158$\mu$m
line intensity rises weakly with $G_0$, for a constant C$^+$ column. Below this
line, the PDR temperature drops below 92~K and the [C~II] 158$\mu$m intensity 
is much more sensitive to $G_0$. The line marked 
$N$(C$^+$)$=10^{21}{x}_{\rm C}\,\rm cm^{-2}$ indicates the boundary between 
a relatively constant column of C$^+$ and a falling column of C$^+$, where 
$x_{\rm C}$ is the fractional abundance of gas phase carbon relative to H 
nuclei. Above and to the
left of the line, the absorption of FUV flux is dominated by grain opacity, and
the column density of C$^+$ is approximately constant, $\sim 10^{21}{x}_{\rm C}\,\rm cm^{-2}$, 
corresponding to all carbon in C$^+$ in a hydrogen nucleus column density of 
$10^{21}\,\rm cm^{-2}$. Below and to the right of this line, H$_2$ and C 
opacity dominate the absorption of FUV flux and $N$(C$^+$) drops significantly below 
$10^{21}{x}_{\rm C}\,\rm cm^{-2}$. 
A hydrogen column density of $10^{21}\,\rm cm^{-2}$ corresponds to an
FUV opacity due to grains of order unity, so this line marks the boundary 
between dust shielding and gas (atomic carbon and H$_2$) shielding of the carbon-ionizing FUV flux.
Below and to the right of this line, which occurs at $G_0/n\simeq 
3\times 10^{-3}\, \rm cm^{-3}$, the column density of C$^+$ is quite sensitive
to the 
ratio $G_0/n$. One of the key reactions which enhances the shielding of the 
FUV by atomic carbon is the neutralization of C$^+$ by neutral and negatively 
charged PAHs. Above and to the left of this line, the column of C$^+$ is set 
by the grain opacity which fixes the C$^+$ column at roughly 
$\sim10^{21}{x}_{\rm C}\,\rm cm^{-2}$. 
Therefore, above this line and in the region where $n>n^{\rm C}_{cr}$ and 
$T>92$~K,
the [C~II] 158$\mu$m line intensity is mainly determined by the C$^+$ column 
which is relatively insensitive to changes in $n$ and $G_0$. 
		
In calculating the  temperature structure and line emission from PDRs, we have
used the TH85 escape probability formalism for a 1-D semi-infinite slab. This
formalism assumes that the IR {\it line} photons
escape from only the front face of the slab, and their average intensity 
is the integrated escaping emissivity (or flux) divided by 2$\pi$. The 
approximation is quite good for [O~I] 63$\mu$m emission, which is optically
thick in the PDR, but only marginally so for the [C~II] 158$\mu$m emission, 
which typically has line center optical depths of $\sim 1$. Therefore, the 
[C~II] and [O~I] line intensities  plotted in the following figures are the 
emergent fluxes from one face of the PDR layer divided by 2$\pi$. 

In Figure 3, we show the intensity of the important PDR cooling transition,
[C~II] 158$\mu$m ($^2P_{3/2}$--$^2P_{1/2}$), as a function of $n$ and $G_0$. 
The behavior may be understood by referring to 
Figure 2 and examining several distinct regions of the parameter space. 
The intensity of the [C~II] line scales approximately
as $I({\rm C~II})\propto N($C$^+)e^{(-92/T)}/[1+(n_{cr}^{\rm C}/n)]$
where $N($C$^+)$ is the column density of ionized carbon, $\Delta E/k\sim
92\rm\,K$ is the energy of the upper state of the 158$\mu$m transition, 
and $n_{cr}^{\rm C}\simeq 3\times 10^3\,\rm cm^{-3}$ is the critical density of
the 158$\mu$m transition. In the lower left portion of the figure, $T<92$~K,
$n<n^{\rm C}_{cr}$, and $N($C$^+)\sim 10^{21}{x}_{\rm C}$ 
since dust dominates the absorption
of carbon-ionizing photons. In this lower left region (roughly $G_0\lesssim 
10$, $n\lesssim 3\times 10^3\,\rm cm^{-3}$), the column of C$^+$ is constant, 
but the intensity of [C~II] 158$\mu$m emission rises very sharply with
increasing  $G_0$ because $T<92$~K and $T$ increases with $G_0$. On the other
hand, for  fixed $G_0$, the [C~II] 158$\mu$m intensity is not very sensitive
to $n$. This is  because $T$ drops with increasing $n$ (see \S 3.1) which
offsets the increasing collision rate (the $n/n^{\rm C}_{cr}$ term). More
insight into this cancellation can be gained by noting that [C~II] 158$\mu$m
emission dominates the cooling in this  regime and that therefore the [C~II]
158$\mu$m intensity is proportional to the total heat input via grain
photoelectric heating. In this regime, the grains  are quite neutral and the
column-integrated grain photoelectric heating is  just proportional to $G_0$
and independent of $n$. Roughly 3\% of the incident  FUV flux is converted to
[C~II] 158$\mu$m flux, with most of the rest emerging as  grain IR continuum.

In the upper left region (roughly $G_0\gtrsim 10$, $n\lesssim 3\times 10^3\,\rm
cm^{-3}$), the column of C$^+$ is constant and the intensity of [C~II]
158$\mu$m  emission is nearly independent of $G_0$, since $T>92$~K and
increases in $T$ do  not appreciably change the intensity. On the other hand,
the intensity now  scales with $n/n^{\rm C}_{cr}$.

In the upper right region of the plot, above the $N($C$^+)\simeq 10^{21}
{x}_{\rm C}
\,\rm cm^{-2}$ line and for $n\gtrsim 3\times 10^3\,\rm cm^{-3}$, the column of
C$^+$ is constant, $T>92$~K and $n>n^{\rm C}_{cr}$, and so the intensity of 
[C~II] 158$\mu$m is fairly constant and only weakly dependent on $G_0$ and $n$
(the intensity changes by $\sim 10$ as $G_0$ and $n$ vary by more than $10^3$).
In this region of parameter space, [O~I] cooling is dominant.

In the lower right region of the plot, below the $N($C$^+)\simeq 10^{21}
x_{\rm C}
\,\rm cm^{-2}$ line, $N($C$^+)$ begins to drop significantly. The crucial 
reaction neutralizing
C$^+$ is not electronic recombination, which is slow (rate coefficient of 
order $10^{-11}\,\rm cm^3\,s^{-1}$ with an electron abundance of order 
$10^{-4}$), but  the neutralization of C$^+$ by neutral and negatively charged 
PAHs (rate coefficients of order $10^{-8}-10^{-6}\,\rm cm^3\, s^{-1}$ with an
abundance of $4\times 10^{-7}$). The neutral carbon reacts with various 
species including OH, H$_2^*$ and O in a set of reactions leading to CO
(c.f. Hollenbach \& Tielens 1999, Fig. 9b.) The column of C$^+$ and the 
temperature are controlled by the ratio $G_0/n$, which is why lines of constant
$G_0/n$ have constant [C~II] 158$\mu$m intensity in this region.

For high values of $G_0$ (and hence high temperatures), the 
conditions become appropriate for strong [O~I] 63$\mu$m emission, which has a
higher upper state energy ($E_{upper}/k\sim $228~K) than that of [C~II]
158$\mu$m (92~K).  Figure 4 shows the ratio of the intensity of the 
[O~I] 63$\mu$m ($^3P_1$--$^3P_2$) transition to that of the [C~II] 
158$\mu$m transition. At high $G_0$ the PDR surface temperature is above the 
upper state energy for both [C~II] and [O~I] and the column of warm \cp~ and 
O is set by dust 
attenuation of the FUV. The ratio [O~I]/[C~II] is therefore roughly constant 
for fixed $n$ in this regime. However, at
low $G_0$ or low $T$, the 63$\mu$m intensity falls faster with decreasing 
$G_0$ than the 158$\mu$m intensity and
the ratio drops. The critical density of the [O~I] 63$\mu$m 
transition
($n_{cr}^{\rm O}\sim 5\times 10^5\,\rm cm^{-3}$) is considerably higher than 
that of the [C~II] transition, so at densities greater than $\sim 3\times 
10^3\,\rm 
cm^{-3}$  the 63$\mu$m intensity continues to increase rapidly with density 
while the 158$\mu$m intensity is relatively constant with increasing density; 
thus for $T\gg 100$~K (or $G_0\gg 10$), and $3\times 10^3\,{\rm
cm^{-3}}<n<5\times 10^5\,\rm cm^{-3}$, 
the [O~I] 63$\mu$m to [C~II] 158$\mu$m intensity ratio climbs rapidly 
with $n$.

Figure 5 shows contours of the [O~I] 145$\mu$m ($^3P_0$--$^3P_1$)/[O~I] 
63$\mu$m
($^3P_1$--$^3P_2$) intensity ratio as a function of $G_0$ and $n$. The [O~I]
63$\mu$m line is generally optically thick (line center optical depths of 
several) over our entire parameter space. The 145$\mu$m line, by contrast, 
is optically thin over the entire parameter space. The ratio varies by less 
than an order of magnitude over nearly the entire range of $G_0$ and $n$. Both 
transitions have similar critical densities so the ratio is relatively 
insensitive 
to density, but does decrease somewhat for $n>10^5\,\rm cm^{-3}$. 
Excitation of the 145$\mu$m transition requires $\Delta E/k$ of 
$\sim 325$~K, about 100~K higher than the 63$\mu$m transition, and so the ratio
becomes more sensitive to temperature (or $G_0$) for $T\lesssim
300$~K or  $G_0\lesssim 10^2$.
 
Figure 6 shows the ratio of the sum of the strengths of the two major cooling
lines, [O~I] 63$\mu$m and [C~II] 158$\mu$m, to the bolometric FIR dust
continuum emission. Care must be taken when using this plot because the
results depend both on the geometry and morphology of the observed region and
on the fraction of dust heating which occurs outside the FUV band. As discussed
above, we assume that IR line photons escape only from the front face of the
PDR and that the intesity in a line is the integrated line flux divided by 2$\pi$. 
In contrast to the atomic line emission, the grains radiate optically-thin
FIR continuum radiation which escapes from the front and back of most
interstellar clouds. Therefore if a finite slab is illuminated from one side
by a flux $G_0$, and all grain heating is by FUV photons, the FIR continuum
intensity radiated back is given  by $I_{FIR}=1.3\times 10^{-4}\,G_0\,\rm
erg\,cm^{-2}\,s^{-1}\,sr^{-1}$ (recall that $G_0=1$ corresponds to $1.6\times
10^{-3}\,\rm erg\,cm^{-2}\,s^{-1}$ and  $1.3\times 10^{-4}=1.6\times
10^{-3}/4\pi$). Typically for OB stars, however, there is about equal heating
of the grains by photons outside the FUV band  (TH85), so that the observed
intensity is roughly $I_{FIR}\simeq 2\cdot 1.3\times 10^{-4}\,G_0\,\rm
erg\,cm^{-2}\,s^{-1}\,sr^{-1}$. Figure 6 uses this approximation for the
emergent infrared intensity. 

For the case of clouds in the active regions of galaxies, or where many 
clouds are contained within the beam, another situation arises. Here the clouds
are illuminated from all sides and the observer sees the optically-thin FIR
continuum from both the near and far side of the cloud, whereas the [O~I] 63$
\mu$m emission (and to a lesser extent the [C~II] 158$\mu$m emission) is only 
observed from the side 
facing the observer. For clouds illuminated on all sides by $G_0$, the observed
([O~I] 63$\mu$m + [C~II] 158$\mu$m)/FIR ratio is simply the ratio of the 
escaping intensities of line photons to grain continuum for each cloud. This 
corresponds to $I$([O~I] 63$\mu$m +[C~II] 158$\mu$m)/$2 I_{FIR}$, 
where 
$I_{FIR}\simeq 2\cdot 1.3\times 10^{-4}\,G_0\,\rm erg\,cm^{-2}\,s^{-1}\,
sr^{-1}$ accounts for an approximate bolometric correction of 2 as before.
Therefore, when modeling clouds illuminated {\it on all sides}, the contours 
of Figure 6 should be reduced by a factor of 2. 

The contours of Figure 6 may be understood by analyzing the effect of $G_0/n$
on photoelectric heating and on the depth to which C$^+$ and O exist in the 
PDR. For $G_0/n\gtrsim 3\times 10^{-3}\,\rm cm^3$, C$^+$ and O exist to a depth
of at least $\tau_{FUV}\gtrsim 1$, where $\tau_{FUV}$ is the FUV optical depth
of the dust.
Since [C~II] 158$\mu$m and [O~I] 63$\mu$m dominate the cooling when they are
abundant, the emission by these species is just equal to the grain 
photoelectric heating of the gas. The grain photoelectric heating efficiency, 
$\epsilon$, is defined as the ratio of the FUV energy which goes into the
gas to the FUV energy which goes into grain heating (TH85). For $n\le 10^5\,
\rm cm^{-3}$ and $G_0/n\gtrsim 10^{-2}\,\rm cm^3$, the contours track this 
efficiency and are given by roughly $\case{1}{2}\epsilon$ where the factor of 
$1/2$
arises because of the bolometric correction. Note that the constant contour 
lines track constant ratios $G_0/n$ since this ratio controls the grain 
positive charge which, in turn, controls the gas heating. The relationship 
breaks down in the upper right quadrant of the figure, where $G_0\gtrsim 10^4$
and $n\gtrsim 10^5\,\rm cm^{-3}$, since the PDR temperature is high ($T\gtrsim
5000\,\rm K$) and the ``recombination cooling'' of the gas caused by electrons
recombining on grains becomes significant. In other words, the energy of the 
recombining electrons becomes comparable to the energy of the photoejected
electrons and the gas heating is significantly reduced. 

For $G_0/n\lesssim 3\times 10^{-3}\,\rm cm^3$, the \cp/C/CO transition is 
drawn to the cloud surface. In this
regime, much of the grain photoelectric heating of the gas emerges as [C~I] 
or CO line emission, and so the ([O~I] 63$\mu$m +[C~II] 158$\mu$m)/FIR ratio is
reduced below  $\case{1}{2}\epsilon$.  

\subsection{[C~I] 370$\mu$m, 609$\mu$m and CO Rotational Lines}

In this section we present results for the strengths of lines which are 
generally produced in gas deeper within PDRs, where ionization of carbon 
gives way to neutral carbon (C$^0$) and carbon monoxide (CO). 
These lines lie at  
millimeter and submillimeter wavelengths and dominate the cooling at moderate
depths in PDRs. For high $G_0/n\gtrsim 10^{-3}$,  these
species exist at depths of $A_V\sim 2-10$. In these PDRs most of 
the FUV heating of the gas occurs at $A_V\lesssim 1$ and the CO and 
[C I] lines are generally weaker than
[C II] and CO. However, for low values of $G_0/n\lesssim 3\times10^{-3}$, 
gas shielding
allows C and CO to dominate the cooling at small column densities 
into the cloud, and the [C I] and mid-$J$ CO intensties exceed those of
[C II] and [O I]. The [C~I] and mid-$J$ CO lines 
are therefore useful for determining
the details of the chemistry and penetration of the incident FUV in PDRs. 
  
In Figure 7, we show the [C~I] 609$\mu$m ($^3P_{1}$--$^3P_{0}$) intensity, 
as a function of $n$ and $G_0$. As discussed
in TH85, Tielens \& Hollenbach (1985b) and HTT91, the column 
density of neutral carbon is insensitive to the strength of the external FUV
field. Only the depth at which the C$^+$/C/CO transition occurs depends on
the FUV field (see TH85 for details). The intensity of [C~I] 609$\mu$m emission
increases slightly
with density up to $n_{cr}^{CI 609}\sim $ 5 $\times 10^2\,\cc$, reflecting the
greater collisional excitation with $n$ for $n<n_{cr}$, but then slowly
{\it decreases} with increasing density, reflecting a smaller column of C
with increasing density. Neutral C is mainly destroyed by FUV ionization and 
is formed 
by electron recombination with \cp, neutralization of \cp~by negatively 
charged PAHs, and FUV photodissociation of carbon-containing molecules. 
At high densities, neutral carbon
may be formed by FUV photodissociation of CH$_2$ and CH$_3$ and the reaction
$\rm H+CH\rightarrow C+H_2$; at 
lower densities, the photodissociation of CH and CH$_2$ also contributes to 
the abundance of C since these molecules cannot quickly react with O to form
CO. Note that the entire dynamic range of the line intensity is less than a 
factor of 20 over most of our $G_0/n$ parameter space.

Figure 8 shows the ratio of the two submillimeter neutral carbon 
transitions, [C~I] 370$\mu$m/[C~I] 609$\mu$m. 
Both transitions have relatively low upper state energies ($E_{609}/k\sim
24$~K, $E_{370}/k\sim 63$~K). Therefore, the ratio has a very weak temperature 
dependence in most of the parameter space since $T$ is generally
$\gtrsim 50$~K in the [C~I] emitting region. In addition, the critical 
densities
of the two transitions are similiar ($n_{cr}^{CI609}\sim 3\times 10^2\,\cc$, 
$n_{cr}^{CI370} \sim 2\times 10^3\,\cc$). 
Since emission in both lines is from the same species, the ratio is unaffected 
by the abundance of neutral carbon. The figure clearly shows that the ratio 
is insensitive to the strength of the radiation field as long as the 
temperature is large enough to excite both transitions; only in the high 
$n$, low $G_0$ region is temperature a factor since this combination leads
to low temperatures at the C$^+$/C/CO transition. 
Below the critical densities, the intensity ratio varies by a factor of two in
one order of magnitude in density, while above the critical densities it
takes nearly three orders of magnitude in density to change the intensity ratio
by a factor of two. 

Figure 9 shows the ratio of intensities in the [C~II] 158$\mu$m and the
CO $J=1-0$ transitions. Note that our PDR model assumes that the cloud is 
sufficiently thick ($A_V>4$) that a C$^+$ to CO 
transition front exists in the cloud. Both transitions have similar critical 
densities (c.f. Table 2). The CO $J=1-0$ line rapidly becomes optically thick
once \cp converts to CO. 
As discussed in Wolfire et al. (1993), the CO $J=1-0$ intensity is very 
insensitive to $G_0$, because the temperature of the CO $J=1-0$ ``photosphere''
is very insensitive to $G_0$. Typically, $T_{\rm CO}\sim 10-50$~K for $G_0\sim
1-10^5$. Only at very small ratios of $G_0/n$ does $T$ drop below 10~K at the
\cp/C/CO transition so that the $e^{-5.5{\rm K}/T}$ term in the collision rate
becomes important and the level populations become subthermal.  [Note that 
Figure 1 shows the surface temperature and not the temperature of
the \cp/C/CO transition.]
The [C~II] 158$\mu$m/CO $J=1-0$ ratio is largely 
determined by the column density of C$^+$ and $T_S$ which drop (c.f. Figure 2) 
as gas shielding begins to dominate.
As mentioned above, the \cp~ column density is roughly set by the 
ratio $G_0/n$ for $G_0/n<3\times 10^{-3}\,\rm cm^3$, and thus
the decrease in the intensity
ratio seen in the lower right portion of the figure is dominated 
by the drop in the \cp~column. In addition, the ratio drops slightly for 
decreasing $G_0$ 
as the temperature of the [C~II] emitting gas drops below 92~K. Note that over
a large part of $G_0/n$ parameter space, corresponding to ratios $G_0/n\simeq
10^{-3}-1\,\icc$, the [C~II]/CO intensity ratio is $10^3-10^4$, a value
often observed in galactic PDRs (Stacey \ea 1991).

Figure 10 shows the ratio of intensities in the [C~I] 609$\mu$m and the 
CO $J=1-0$ transitions. The behavior is similar to that in the previous
figure. For densities below $n_{cr}\sim 10^3\,\cc$, both transitions increase
in strength with increasing $n$. However, CO increases faster with $n$ than 
[C~I], so the [C~I]/CO ratio drops with $n$; for [C~I] the increase in 
collisional 
excitation is somewhat offset by a decrease in [C~I] column, as described 
above.
Above $n^{{\rm CO}\,J=1-0}_{cr}$, the [C~I] strength 
decreases as the column density decreases with $n$. 
The CO intensity meanwhile is approximately
constant for $n>10^4\,\cc$, so the intensity ratio falls.

Figure 11 shows the intensity of the CO $J=1-0$ 2.6mm transition as a function
 of $G_0$ and $n$. For densities below the critical density 
($n_{cr}^{{\rm CO}\, J=1-0} \sim 5\times 10^3 \cc$), the line intensity 
increases roughly linearly with $n$; above the critical 
density, the line intensity changes far more slowly. Since this line is 
generally optically thick, the intensity is mainly sensitive to the temperature
at the surface of the C$^+$/C/CO transition region, especially for 
$n>n_{cr}^{{\rm CO} \,J=1-0}$. Since the upper state energy 
lies only 5.5~K above ground, the temperature sensitivity increases for very
low values of $T_{\rm CO}$; this may be seen in the lower right hand side of
the figure where the intensity is sensitive to increases in $G_0$.
Again it should be emphasized that the temperature, $T_{\rm CO}$, at which CO 
goes optically thick is an extremely weak function of $G_0$. Often this 
temperature is roughly the grain temperature at the \cp/C/CO boundary. 
The grain temperature varies as 
$G_0^{0.2}$ at the cloud surface, but has an even weaker dependence on 
$G_0$ at the \cp/C/CO boundary since the boundary moves to greater depths as 
$G_0$ increases. 
 
Figure 12 shows the CO $J=2-1/J=1-0$ intensity ratio. In general, both lines
are optically thick in PDRs and the $J=2-1$ line becomes optically thick closer
to the PDR surface (i.e., at higher $T$ than the $J=1-0$ line). The ratio is 
only weakly dependent on gas density over most of the parameter space. Both 
transitions have similar critical densities (c.f. Table 2)
and the intensity ratio 
increases only slightly with $n$ above this value. For very low
values of $G_0/n$ the temperature at the \cp/C/CO transition falls below the
energy of the $J=2-1$ transition upper state and the ratio drops. The intensity 
ratio in Figure 12 may be converted to an integrated antenna temperature ratio, 
$\int T_A[2-1]dv/\int T_A[1-0]dv$, by multiplying by 0.125; this conversion is valid
if one assumes that both transitions have the same intrinsic line width and fill 
the same fraction of the beam. The behavior of the contours in Figure 12 
can also be seen in Figures 13-15, where the ratio of the $J=3-2$,
$J=6-5$ and $J=15-14$ intensities to that of the $J=1-0$ transition are 
plotted (the intensity ratio in Figure 13 can be converted to an antenna temperature
ratio by multiplying by 0.037). 
All of these figures show the same behavior as that seen in Figure 12,
except that the critical densities become increasingly
larger so that the density dependence of the ratio is strongest for the 
highest-$J$ transitions. In addition, at $n\ll n_{cr}$, the relative 
intensities reflect the collisional excitation rates out of the $J=0$ ground 
state. As $J_{upper}$ increases, the rates get progressively smaller relative
to the rate to $J_{upper}=1$. Note particularly that in Figure 15, the 
$J=15-14$
transition never gets nearly as strong as the $J=1-0$ transition except at
very high densities, $n\sim 10^{5.5}\,\cc$. For this line, which lies $\sim 
600$~K above ground, there is both a temperature and density sensitivity
which makes the $J=15-14$/$J=1-0$ ratio extremely high in the upper right hand quadrant of the 
figure. The strength of the $J=15-14$ transition drops so rapidly below and 
to the left of the plotted contours that the ratio of the two lines falls
to $\lesssim 10^{-4}$ as either the temperature or density fall;
note that the strength of the $J=1-0$ line is relatively constant over the 
entire parameter space, varying by only $\sim$25 over the full range of $n$ 
and $G_0$ (c.f. Fig.11). 

\subsection{Effects of Varying Metallicity, $A_V$ and Cloud Geometry}
As noted in the introduction, many galaxies show evidence for low metallicity,
and the Milky Way Galaxy has significant metallicity gradients, 
and so we are motivated to explore the effects of lowered gas phase
abundances and lowered dust fractions on the results of our models.
In addition, we study the effect of varying the total column density or
$A_V$ through an irradiated cloud.
We explore results for both planar and spherical PDR clouds with reduced 
gas metallicity and dust fractions. 

In this section, we focus on emission from two
PDR diagnostics: the [C~II] 158$\mu$m transition, which is emitted primarily 
from the warm atomic regions at $A_V\lesssim 1$; and the CO $J=1-0$ transition,
which is generally emitted from cooler molecular regions at $A_V\gtrsim 2$. We
chose 
these two transitions since they are bright emission lines and are readily 
seen in a wide variety of sources (e.g. Stacey \ea 1991). 
Comparable
resolution observations ($\sim 1\arcmin$) in both transitions have existed for
several years
for Galactic sources, normal and starburst galaxies, and low-metallicity 
galaxies. 
Our results show that the ratio of the two transitions is sensitive to the 
total PDR visual extinction, since CO line emission comes from higher $A_V$. 
This effect can be exaggerated in spherical PDR models where the volume of 
[C~II] emitting gas may be much larger than that of the CO emitting gas.
Lowering the metallicity enhances this effect for clouds of fixed hydrogen 
column, $N$.  
    
\subsubsection{Effects of Varying Metallicity and $A_V$}
For our planar PDR models, we have studied our entire $n-G_0$ parameter space
with metallicities down to $Z=0.1$. We take a reduced metallicity to
correspond to a reduction in both the gas phase heavy element abundance 
and the dust abundance. Therefore,
a metallicity $Z=0.1$ corresponds to all of the gas phase abundances in Table~1
falling to 10\% of their standard values; in addition, the dust abundance,
PAH abundance, corresponding visual extinction cross section per H atom,
and H$_2$ formation rate coefficient $R_{form}$ drop to 10\% of their standard
value. Since our standard PDR models are calculated to a visual extinction of
$A_V=10$, the effect of lowering the dust abundance is to increase the
physical extent  and total hydrogen column density, $N$ of
a PDR model. A PDR with visual extinction $A_V$ and metallicity 
$Z$ has a physical size $L=N/n=2\times 10^{21} A_V/(Zn)\,\rm cm$. Therefore, 
for a 
metallicity of $Z$=0.1 relative to $Z$=1, we increase the size $L$ by a factor
10
and the hydrogen column density from $2\times 
10^{22}\rm\, cm^{-2}$ to $2\times 10^{23}\rm\, cm^{-2}$ in order that the 
total $A_V=10.$

In Figures 16, 17 and 18 we show the relative strengths of the CO $J=1-0$ and 
[C~II] 158$\mu$m lines for models with densities of $10^3$ and $10^5\,\rm 
cm^{-3}$, 
for all $G_0$, and metallicities of $Z=1.0$ and $Z=0.1$. Figure 16 shows
results for models with total visual extinction $A_V=10$. Figure 17 has 
$A_V=3$, 
and Figure 18 shows results for $A_V=1.$ In producing these results, we have
explicitly calculated the structure of 1-D PDR models with lowered metallicity.
However, the low $A_V$ results are 1-D PDR models truncated at $A_V=1$ and 
$A_V=3$ from the surface.
The strengths of the two emission lines are shown relative to the
strength of the FIR continuum. We apply a correction to the FIR emission
expected from the low $A_V$ models such that only that portion of the FUV flux
which can be absorbed in the reduced column is reemitted as FIR continuum.
 We again include the factor of two bolometric 
correction to the strength of the FIR field due to illumination outside the 
FUV band
(c.f. \S3.2). Our results illustrate that the effects of 
metallicity are minor for the parameters of interest. To first order, changing
the metallicity by a factor 10 while keeping $A_V$ constant has little effect
since there is an approximate scaling. Since the species responsible for
the heating (small grains) and the species responsible for the cooling (carbon
and oxygen) are reduced by the same factor $Z$, the temperature in the warm 
PDR zone
remains approximately constant while changing $Z$. In addition, the column of
coolant species like O and \cp~ in the warm zone also remain constant for high
$G_0/n\gtrsim 3\times 10^{-3}\,\rm cm^3$,  although the H column increases.
Therefore, results for models with the same total $A_V$ should be similar. 

There are, however,  some chemical processes which affect the [C~II]/CO ratio
as $Z$ is varied. 
Recall that a prime destruction mechanism for \cp~ at low $G_0/n$ is by 
collisions with 
neutral or negatively charged PAHs. The total PAH abundance as well as the 
fraction of PAHs 
which are negatively charged has a $Z$ dependence, with PAH and 
PAH$^-$ increasing with $Z$ (more gas phase electrons result in more neutral 
or negatively 
charged PAHs). Therefore, the C$^+$ column increases 
with decreasing $Z$ for $G_0/n\lesssim 10^{-3}\,\rm cm^3$. Figures 17 and 18 
show this effect,
[C~II] increasing with decreasing $Z$, for $n\sim10^5\,\rm cm^{-3}$ and 
$G_0\lesssim 100$.  At lower
densities and higher FUV field strengths, dust shielding dominates for 
both metallicities, the temperature exceeds 92~K so the \cp~ intensity is 
insensitive
to $T$, the C$^+$ column is relatively constant, and the curves converge. 

For high values of $G_0/n\gtrsim 3\times 10^{-3}\,\rm cm^3$, Figures 17 and 18
show the effect of
photodissociation of CO when $A_V=3$ and $A_V=1$, respectively. For high 
values of $G_0/n$ 
and the relatively low total slab opacity of $A_V=1$, FUV radiation nearly 
completely 
photodissociates
the CO throughout the PDR and the CO $J=1-0$ line becomes optically thin and 
weak. 
Carbon is ionized throughout nearly the entire slab. Neither gas shielding
nor dust shielding is sufficient to create a \cp/C/CO transition. 
Note that the case of $Z=0.1$ and $A_V=1$ presented in Figure 18
corresponds to the same {\it total} column of hydrogen nuclei as the case 
$Z=1.0$ and $A_V=10$
presented in Figure 16. As can be seen by comparing Figures 16, 17 and 18, 
the [C~II]/CO ratio is vastly different 
for $G_0\gtrsim 10$, because of the steep decline in the CO intensity for 
$A_V\ll 10$. 
Thus, reducing $Z$ in a cloud of given {\it total column
density} has a dramatic effect, while reducing $Z$ in a cloud of given ${\it 
A_V}$ has much less effect. Note that once $A_V\lesssim 1$, not all of the
incident
radiation is absorbed by the slab and reradiated, so that $I$(FIR) is reduced
compared to higher opacity cases, an effect which we have included in Figures
17 and 18. 

In Figure 16, we also indicate the trend in the [C~II]/CO ratio for the case
of unresolved {\it spherical} clouds illuminated from the outside so that the 
outer layers of
the clouds have carbon mostly in \cp, while the central regions contain carbon
in CO. The arrow extending to lower CO intensity indicates that for spherical 
clouds,
the emitting area of \cp~becomes larger than that of CO. For clouds with 
visual extinctions to their centers of $A_V=10$ and constant hydrogen density 
from the 
outer \cp~zone to the inner CO core ($n_{\rm C^+}=n_{\rm CO}$), this effect is
 minor. The 
\cp/C/CO transition occurs at $A_V\lesssim 3$ into the cloud, and therefore the
 CO 
photosphere has almost the same radius as the entire cloud (the \cp~ zone can 
be considered
a thin shell on the outside of this spherical cloud).
Lowering the total $A_V$ of the cloud, however, tends to move
the CO photosphere in towards the center of the cloud, reducing the emitting
area of the optically thick CO emission region and enhancing [C~II] relative 
to CO. The same effect occurs if the hydrogen density of the CO core, $n_{\rm
CO}$ is higher than 
that of the [C~II] emitting region, $n_{\rm CII}$: a higher density CO core 
has a smaller 
physical size for a given $A_V$ and so [C~II]/CO can be enhanced by
a large factor by increasing $n_{\rm CO}$ relative to $n_{\rm CII}$. 
The arrow in Figure 16 shows the trend with an increase in the CO core density
or a decrease 
in $A_V$ for spherical clouds. We emphasize that this effect applies to 
unresolved cores.
    
\subsubsection{The CO/H$_2$ Ratio in Spherical Clouds}
The CO $J=1-0$ luminosity, $L[{\rm CO}\, J=1-0]$, is used as a tracer of $\rm
H_2$ mass, $M_{\rm H_2}$.  Here we investigate the effects of total cloud
column density $N$ and metallicity $Z$ on the ratio $L[{\rm CO}\,J=1-0]/M_{\rm
H_2}$. As noted in the previous  section, the product $NZ$ or the $A_V$
through the cloud has the largest effect on the CO emission from a PDR model.
We also probe the sensitivity of $L[{\rm CO}\, J=1-0]/M_{\rm H_2}$ to
variations in $G_0$ and $n$. We have made a very simplified calculation 
of the ratio
$L[{\rm CO}\,J=1-0]/M_{\rm H_2}$  to be expected from 
spherical clouds by assuming that our 1-D slab model results 
accurately  describe the (escaping) volume emissivity
$j(N)$ in the $J=1-0$ line as a function of  $A_V$ for a spherical cloud. 
We have not explicitly calculated spherical models nor
have we included an escape probability appropriate for spherical geometry. To
find the CO luminosity of a particular cloud we calculate \begin{equation}
L[{\rm CO}\, J=1-0]=\int 4\pi j(N) r^2 dr
\end{equation}
where $r=[N_{total}-N]/n$ is the radial coordinate and $j(N)$ is the
emissivity in the 1--0 line at a column $N$ from the surface computed from our
slab models. For example, a
cloud with a column density of $2\times 10^{21}\,\rm cm^{-2}$ from the surface
to the center has the emissivity  calculated for $A_V\sim 0$ at the surface
and the emissivity calculated for $A_V=1$ at the center. 
We calculate $L[{\rm CO}\,J=1-0]$ as a function 
of the column density to the center of a spherical cloud for columns 
up to $10^{22.3}\,\rm cm^{-2}$ for the 
case of $Z=1$ and for columns up $10^{23.3}\,\rm cm^{-2}$ for the case of
$Z=0.1$. For comparison, St\"orzer \ea (1996) have explicitly calculated
the structure of spherical PDRs; they find that the temperature and chemical 
profiles of constant density spherical PDRs are very similar to those for
plane-parallel PDRs, especially when the column density is high enough that
the C/\cp/CO transition occurs near the cloud surface. They do find that 
this transition occurs nearer the PDR surface for plane-parallel models than 
for spherical models in low column density cases, an effect which is smaller
than 20\% for the cases they present. Since the CO line is generally optically
thick and is relatively insensitive to temperature, we expect
that our simplified calculation will give results within a factor
of 1.5 larger than a detailed spherical model. 

Our results are shown in Figure 19 for $Z=1$ and in 
Figure 20 for $Z=0.1.$ Moving to the right along the x-axis in these figures
is equivalent to increasing the total visual extinction (labelled on top) or
total cloud hydrogen column density (labelled on bottom) to the 
center of the spherical cloud. Results are shown for $G_0=1,10^{1.5}$ and 
$10^{3.5},$ and for densities $n=10^2,\, 10^3$ and $10^4\,\rm cm^{-3}$. 
At very low visual extinctions, the ratio is small since the CO is 
photodissociated
more effectively than the $\rm H_2$. As we move to higher total visual 
extinction, 
the ratio $L[{\rm CO}\,J=1-0]/M_{\rm H_2}$ drops near $A_V
\sim 1$ (for high ratios of $G_0/n$ where dust controls the $\rm H/H_2$
dissociation front)
as molecular hydrogen begins to form but CO is still photodissociated. 
As CO becomes the dominant carbon
bearing species ($A_V\gtrsim 2$), the ratio again begins to rise. In the case
of low  $G_0$, the ratio eventually drops as we move to higher total $A_V$
since CO is optically thick near the surface and $L[{\rm CO}\, J=1-0]$ roughly
scales with cloud area (or $A_V^2$ for fixed $n$) while $M_{\rm H_2}$ scales
with volume (or $A_V^3$ for fixed $n$). Therefore the ratio of the two
declines as $A_V$ (and cloud radius) gets larger. In the case of very high FUV
field  ($G_0\gtrsim 100$), the CO photospheric radius is appreciably smaller
than the $\rm H_2$ radius even when $A_V=10$. The CO photospheric area
therefore increases faster than the  $\rm H_2$ volume and the ratio continues
to rise as $A_V$ increases from 3 to 10. Lowering the metallicity to $Z=0.1$
but fixing  $A_V$ leads to a smaller ratio since the volume (mass) of the
molecular gas  is increased relative to the area of the CO photosphere (c.f.
Figure 20).  The ratio is also a 
strong function of density between $n=10^2\,\rm cm^{-3}$ and $n=10^4\,\rm
cm^{-3}$, due in part to chemical effects and in part to the fact that 
the critical density for this transition is $n\sim 3\times 10^3
\,\rm cm^{-3}$ (c.f. Table 2). 
 
The main points illustrated in Figures 19 and 20 are the following. The 
$L$(CO $J=1-0$)/$M$(H$_2$) ratio varies over many orders of magnitude for
clouds of low $A_V\lesssim 3$. Even if all clouds have the same (high) value
$A_V \approx 10$, there are still order of magnitude changes in the ratio as
$Z$ and  $n$ vary. The drop of $\sim$10 in the ratio as $Z$ drops from 1 to
0.1 is  primarily due to the increased size of the clouds, which changes the
volume ($\sim$H$_2$ mass) by a factor of 10 more than it changes the CO
photospheric area ($\sim$$L$[CO]). For $A_V=10$, a drop in $Z$ from 1 to 0.1
increases the radius of a cloud by a factor 10, increasing the
H$_2$ mass by a  factor of $10^3$ but only increasing the area of the CO
photosphere by a factor of $10^2$. For virialized clouds, as the ratio $M/R$
increases by a factor of 100, the turbulent Doppler velocity $\delta_{v_D}$ 
should also increase by a factor 10. Wolfire, Hollenbach \& Tielens (1989) 
showed that variations of $\delta_{v_D}$ by a factor 2 had little effect on 
the line intensities, but this larger 
increase in  velocity dispersion might have a significant effect on the
intensity of the CO $J=1-0$ line; we have not included this effect in our 
models. The $L$(CO $J=1-0$)/$M$(H$_2$) ratio also
drops by about an order of magnitude for fixed $Z$ as $n$ decreases from
$10^4\,\rm cm^{-3}$ to $10^2\,\rm cm^{-3}$, primarily due to subthermal
excitation of CO. However, the ratio is quite insensitive to large variations
(over 3 orders of magnitude) in $G_0$. The standard ratio  $L$(CO
$J=1-0$)/$M$(H$_2$) is typically taken as $\sim 8\times 10^{-6}L_{\odot}
/M_{\odot}$ (e.g. Young et al. 1986)  which corresponds to $Z$=1, $A_V$=10,
$n\sim 10^4\,\rm cm^{-3}$ and $G_0\sim 1-100$ (c.f. Figure 19). 

\subsection{Application to Extragalactic Sources}

\subsubsection{Sources smaller than the beam}
In the previous sections of this paper, we have implicitly assumed that a 
single PDR is contained within the observer's beam. This is the case, for 
example, for the Orion PDR behind the Trapezium where a typical 60\arcsec~
beam covers $\sim$ 0.15 pc, a size much smaller than the emitting region of
$\sim$ 1 pc. In extragalactic observations, however, the physical size of 
the beam at the distance of the source (typically $\sim$kpc)
is large compared with a single PDR
and the observed emission arises from many separate PDR surfaces. In this 
subsection, we mention several cautionary notes when applying our theoretical 
models to these types of observations and further outline the additional 
analysis which can be carried out for extragalactic sources. 

First note that the observed line width measures the dispersion velocity of 
PDRs (clouds) within the beam rather than the turbulent gas velocity, 
$\delta v_D$,
within a single PDR. Thus, the appropriate $\delta v_D$ to use for 
extragalactic 
observations is of the order a few $\rm km\,s^{-1}$ as adopted in this paper
and $\delta v_D$ is not the much larger velocity width of the emission line.

Second, the observed peak line intensity depends on the intensity emitted by
each cloud and the beam filling factor at the central velocity. It is 
important, therefore, to correct as best as possible for the different area 
filling factors for various species. For example, diffuse [C~II] emission may 
fill the beam whereas [O~I] emission may arise from a smaller, high $n$ and 
high $T$ region. In such a case it would not be appropriate to take the ratio
of observed fluxes {\cal F}([C~II])/{\cal F}([O~I]) for use in the contour 
plots presented in  this paper. The ratio of intensities 
{\it I}([C~II])/{\it I}([O~I]) = {\cal F([C~II]})/$\Omega_{\rm [CII]}$
/{\cal F}([O~I])/$\Omega_{\rm [OI]}$ should be used for the ratio of fluxes 
corrected for the separate [C~II] and [O~I] emitting solid angles,
$\Omega_{\rm [CII]}$ and $\Omega_{\rm [OI]}$.  The density, temperature, and
incident FUV field derived from  the observations are necessarily an ensemble
average of the PDRs in the beam.  WTH90 analyzed the bias in the derived
quantities and found that the method is somewhat biased toward the high
density, high $G_0$ regions. 

A third caution is that some portion of the diagnostic line emission, in 
particular
that of [C~II], may arise from non-PDR gas such as from low density HII 
regions or
a diffuse warm ionized component. This emission should be subtracted before 
using the
contour plots. One method for estimating this emission is to rely on HII
region 
modeling as presented, for example, in Carral \ea (1994), or from CLOUDY 
(Ferland 1996) models. Carral \ea estimated that 30\% of
the [C~II] emission in NGC 253 originates in HII regions. Another method may 
be used when
[N~II] emission is observed, as the [C~II] emission from low density HII 
regions scales 
with [N~II] (Heiles 1994).

Having obtained the average density $n$, temperature $T$, and incident FUV 
field $G_0$,
from the far-infrared line and continuum observations, the analysis explained 
in WTH90 
and Carral \ea (1994) can be carried out to determine additional 
characteristics of the
interstellar medium. The first step is to determine the mass of neutral PDR 
gas from the 
observed [C~II] line intensity and the derived gas temperature and density from
observations of [C~II], [O~I] and the FIR continuum flux from the warm dust in 
the PDRs. If CO 
observations are available, then estimates can be made of the molecular gas 
mass, the area
and volume filling factor of the clouds, and the number and radii of clouds. 
The last 
step is to evaluate the source of FUV radiation as being consistent with a 
distribution 
of stars or a central source, and whether the absorbing clouds are correlated 
in space with the FUV source(s). 

We will not repeat the steps here and the reader should refer to the 
original papers for
details; however, a few points should be noted in applying the updated 
models. Note that
the [C~II] cooling plot shown in WTH90 (Fig. 4) is acceptable for use 
with the current
models with the reduced gas phase carbon abundance. This is because the
 plot shows the
cooling per \cp~ atom as opposed to the cooling per hydrogen nucleus and 
thus it remains
unchanged for present models. Also note that only the surface temperature 
is presented 
in this paper, while both a surface temperature and maximum temperature 
were shown in 
WTH90. With the updated PAH heating we find that the maximum temperature is 
at, or close, 
to the surface and we can take $T_S$ to be the characteristic PDR temperature.
The remaining analysis should proceed unchanged from the former papers.     
 
\subsubsection{Application to M82}

Our model results may be used to determine a variety of parameters of the 
ISM in external galaxies. In order to compare our results with those of WTH90,
we have repeated the analysis of the conditions in the starburst galaxy 
M82 using the data from Lord et al. (1995), who measured the strength of the 
[O~I] 63$\mu$m line from the central 
regions of M82 and compared it with the strengths of the 
[C~II] 158$\mu$m line measured by Stacey et al. (1991), the [O~I] 
145$\mu$m line measured by Lugten et al.
(1986) and the FIR continuum emission measured by Telesco \&
Harper (1980). The measured fluxes
in the [O~I] 63$\mu$m, [O~I] 145$\mu$m, [C~II] 158 $\mu$m lines and the 
FIR continuum are 1.31$\times 10^{-10}$, 0.84$\times 10^{-11}$, 1.42$\times
10^{-10}$ and 9.3$\times 10^{-8}$ erg cm$^{-2}$ s$^{-1}$, respectively. 
The [C~II] emission from M82 emerges from a far more extended region than 
the [O~I] emission, and was measured with a larger beam, so we have compared
the line ratios in two limits. At a distance of 3.25 Mpc, the [C~II] and
[O~I] beam sizes are $\sim$1 kpc and 400$\times$200 pc, respectively, so we
are  studying gas in the very central regions of M82.  

In the first case, we assume that the [C~II]
emission has equal intensity over the entire [C~II] beam, so that the [C~II]
flux emerging from the [O~I] emitting region is just the [C~II] flux reduced
by the ratio $\Omega_{\rm [OI]}/\Omega_{\rm [CII]}\sim 0.112$, where 
$\Omega_{\rm [OI]}$ and $\Omega_{\rm [CII]}$ are the beam solid angles of
emission in each of the lines. 
This gives a [C~II] flux from the [O~I] emitting region of 1.59$\times 10^{-11}
$ erg cm$^{-2}$ s$^{-1}$. The resultant ratios are: $F$[O~I] 63$\mu$m/$F$[C~II]
158$\mu$m = 8.24, $F$[O~I] 145$\mu$m/$F$[O~I] 63$\mu$m = 0.064 and 
($F$[O~I] 145$\mu$m + $F$[C~II] 158$\mu$m)/$F$(FIR) = 1.58$\times 10^{-3}$
(which we divide by 2 for an extragalactic source, as described in \S3.2). 
Comparing the first two line ratios and overlaying Figures 4 and 5, we find
typical conditions of $n=10^4\,\rm cm^{-3}$ and $G_0=10^{3.5}$; by comparing
$F$[O~I] 63$\mu$m/$F$[C~II] 158$\mu$m with the line-to-continuum ratio, and 
using Figures 4 and 6, we find $n=10^{3.9}\,\rm cm^{-3}$ and $G_0=10^{3.4}$.
As an additional check on this procedure, we can compare the ratio of the two
neutral carbon transitions (Figure 8), $I$([C~I] 370$\mu$m)/$I$([C~I] 
609$\mu$m)=4.3 (Stutzki et al. 1997) with the far-infrared line-to-continuum 
ratio and again find $n\sim 10^4\,\rm cm^{-3}$, $G_0\sim 10^4$.
For the purposes of the following discussion, then, we take $n=10^4\,\rm 
cm^{-3}$ and $G_0=10^{3.5}$ as representitive conditions in the central region
of M82. Reference to Figure 1 indicates a surface temperature $T_S\sim 400~K$. 
These values of $T_S$, $n$, and $G_0$ are somewhat different than 
those calculated by Lord et al. for the central regions of M82: 
$T_S$ = 230 K, $n$ = 10$^{3.9}$ cm$^{-3}$, and $G_0$ = 10$^{3.0}$, 
due to the changes in our PDR model as described above. 
 
Following the procedure outlined in WTH90, we can estimate the mass of the 
neutral PDR gas from the observed [C~II] flux from the [O~I] emitting
region. To facilitate our comparison with the results of Lord et al. (1995), 
we use a distance of 3.25 Mpc to M82, but we use a revised elemental C
abundance of 1.4$\times 10^{-4}$. Using Figure 4 of WTH90 and a surface 
temperature $T_S=400$ K, we find a cooling power of 
$1.1\times 10^{-21}$ erg s$^{-1}$ cm$^{-2}$ sr$^{-1}$ per \cp~ atom; then eq. 2
of WTH90 gives an atomic gas mass of $M_a=8.83\times 10^6\,M_{\odot}$. 
We can compare this to the total 
molecular mass $M_m\sim 2\times 10^8\,\rm M_{\odot}$ as determined by 
observations of optically thick CO $J=1-0$ emission (Wild et al. 1992). 

Following the method of WTH90, we find an area filling factor of clouds 
$\Phi_A=2.7$ by taking the ratio of the observed [C~II] intensity to the 
predicted intensity as given by Figure 3 for $n=10^4\,\rm cm^{-3}$ and 
$G_0=10^{3.5}$. For comparison, the same procedure, when applied to the [C~I]
609 $\mu$m and CO $J=1-0$ lines, where $I$([C~I] 609$\mu$m) and $I$(CO
$J=1-0$) equal $\sim 10^{-5}$ and $\sim 2\times 10^{-7}$ erg cm$^{-2}$ s$^{-1}$
sr$^{-1}$, respectively (e.g. Stacey et al. 1993; White et al. 1994), also 
yield filling factors of $\sim 2$, according to Figures 7 and 11. 
As discussed by WTH90, for clouds illuminated by an FUV 
radiation field over 4$\pi$ sr, the projected cloud area filling factor
$\Phi_{A}^{\prime}=\Phi_A/4=0.67$. Using the molecular mass estimate, and 
assuming
a molecular-to-atomic gas density ratio $\eta = 3$, we derive a volume filling
factor $\Phi_V=6.2\times 10^{-3}$ (eq. 12, WTH90). With these parameters, we
may calculate the following average cloud properties for the central regions
of M82: cloud radius $r_{cl}=1.7$ pc, number of clouds $N_{cl}\sim 7500$, 
and a mean free path between clouds $\lambda = 365$ pc, assuming a spherical
distribution. Our results
imply cloud masses $M_{cl}\sim 3\times 10^4$ M$_{\odot}$, somewhat smaller 
than massive GMCs in the Galaxy. However, these clouds have thermal pressures
of order $\sim 100$ times those of GMCs, resulting from a factor $\sim 10$  
higher temperature and from a factor $\sim 10$ higher $n$.

For comparison, we have repeated the above calculation assuming that all of
the [C~II] emission from M82 arises from the same gas 
which is responsible for the [O~I] emission (i.e. a much lower [O~I]/[C~II]
ratio). Using the same procedure, we derive the following average properties:
$n\sim 10^{2.7}$, $G_0\sim 10^{2.5}$, $M_a\sim 4\times 10^8$ M$_{\odot}$, 
$\Phi_A\sim 50$, $r_{cl}\sim 23$ pc, $N_{cl}\sim 770$, and $M_{cl}\sim
1.2\times 10^6$ M$_{\odot}$. This large variation in our results for 
cloud masses, numbers of clouds, etc.  determined
above shows the sensitivity of our results to the assumed details of the 
emitting region. Therefore, independent measures which can further constrain the 
conditions in M82 (e.g. SOFIA observations of the spatial extent of the CII emitting
region region) will decrease the range of possible conditions determined from these
models. In the absence of other independent measurements, these models can show trends in
$n$, $T$, pressure and UV field for a large statistical sample of galaxies. These trends
might be compared, e.g. with star formation activity or the presence of AGN.
Our method is more useful for comparing
galaxies than for determining absolute numbers for these properties. 

\subsubsection{Application to NGC 278}

As a second application of our results, we use a similar method to determine 
the average properties of the interstellar medium in the spiral galaxy NGC 278
\footnote{These data were obtained with ISO under the NASA Key Project on
Normal Galaxies (Helou et al. 1996) and will appear in Malhotra et al.
(1999)}.  NGC 278 was observed in a number 
of lines including [C~II] 158$\mu$m, [O~I] 63, 145$\mu$m and [N~II] 122$\mu$m,
with fluxes of 0.69, 0.31, 0.013 and 0.030 $\times 10^{-18}$ W cm$^{-2}$, 
respectively. The measurement of the strength of the [N~II] line, which arises
from ionized gas, gives us an 
independent measure of the contribution of HII regions to the [C~II] line
flux. From theoretical considerations, Heiles (1994) concludes that the [C~II] 
flux from HII regions is $\sim 6.5$ times that of [N~II] 122$\mu$m, 
assuming an HII
region temperature of $10^4$~K and our standard carbon abundance 
(c.f. Table 1). Measurements from the COBE satellite (Wright \ea 1991; 
Petuchowski \& Bennett 1993) show this 
ratio is $\sim 8.7$ for Galactic HII regions. Thus we expect that 7-8 times
the [N~II] flux emerges from HII regions as [C~II] flux. Applying this 
correction, we find that the [C~II] flux from PDR gas in NGC 278 is $\simeq 
0.44
\times 10^{-18}$ W cm$^{-2}$, assuming NGC 278 has the same C/N ratio as the 
Milky Way. IRAS measurements of the 60 and 100 $\mu$m flux
from NGC 278 (Moshir \ea 1990) translate to a FIR flux of $\sim 1.34\times 10^{-16}$ W cm$^{-2}$.

For the measured PDR line strengths, we find [O~I] 63$\mu$m/[C~II] 158$\mu$m,
[O~I] 145$\mu$m/[O~I] 63$\mu$m and ([O~I] 63$\mu$m +[C~II] 158$\mu$m)/FIR 
ratios of 0.69, 0.042 and $5.6\times 10^{-3}$, respectively. Using Figures 4-6
we may estimate the average properties of the ISM in NGC 278. From the pair of 
line ratios, we find $n\sim 5\times 10^3$ cm$^{-3}$ and $G_0\sim 30$. 
From the [O~I]/[C~II]
ratio and the line to continuum ratio (including the factor of 2 correction 
when using Figure 6 for extragalactic sources) we find $n\sim
10^{3}$ cm$^{-3}$ and somewhat higher $G_0\sim 140$. As mean properties, 
then, we assume $n=2\times 10^3$ cm$^{-3}$ and $G_0=70$. We find lower values 
of $G_0$ and $n$ for this galaxy than the values for M82. This is expected 
since 
NGC 278 is a ``normal'' galaxy compared with the starburst nucleus of M82.
Since NGC 278 lies at a distance of 12.4 Mpc (Braine et al. 1993), the 
observations are sampling nearly 4 kpc of the disk of NGC 278, as opposed to 
roughly 1 kpc of the M82 disk, so the sampled 
regions represent a more global average than those of the central regions of 
M82. 

With these average properties, we have repeated the WTH90 process for 
determining
typical parameters of the ISM in NGC 278. From Figure 1, we estimate a surface
temperature $T_S\sim 200$~K, which leads to an estimate of the mass of atomic
gas from PDRs, $M_a\sim 5\times 10^7\,M_{\odot}$, more than 5 times larger 
than in M82. This can be compared with the estimated mass of molecular gas,
$M_m\sim 1.7\times 10^9\,M_{\odot}$ (Young et al. 1996). Following the WTH90
method,  we find an area filling factor of clouds $\Phi_A=0.35$. Using the
same estimate of the molecular-to-atomic gas density ratio, $\eta=3$, we find
the number of molecular clouds, $N_{cl}\sim 3\times 10^4$, cloud radii,
$r_{cl}\sim 5.5 $~pc, and cloud  mass, $M_{cl}\sim 3.5\times 10^4\,M_{\odot}$.
Assuming the GMCs are confined to a galactic disk with thickness $\sim 200$ pc,
we find a volume filling factor in the disk of $\simeq 3.5\times 10^{-3}$ and
a mean free path between clouds in the disk of $\sim 1\,\rm kpc$. We note that
compared with our first model of M82, the clouds have similar mass and size.
The mean free path between clouds is also similar, although the clouds in 
M82 are a bit smaller in size and closer together as might be expected. 
However, the clouds in M82 have much greater incident FUV fluxes and are at
much higher thermal pressures. This, too, is not entirely unexpected.   
  
\subsubsection{Application to the LMC}

As an application of the model results presented in Figures 16-20, we
compare the [C~II]/FIR and CO/FIR ratios for the LMC. Poglitsch et al. (1995)
observed 30 Doradus in the LMC with a 1$\arcmin$ beam ($\sim$ 15 pc at the LMC)
and found [C~II] 158
$\mu$m, CO $J=1-0$ and FIR continuum intensities of $1\times 10^{-3}$, 
$1.4\times 10^{-8}$ and 0.7 erg s$^{-1}$ cm$^{-2}$ sr$^{-1}$, respectively.
These results give $I$[C~II]/$I$(FIR)=$1.4\times 10^{-3}$ and $I$[CO]/$I$(FIR)
=$2.0\times 10^{-8}$. These values are indicated with a large `{\bf X}' in
Figures 16 and 17. The figures show that the observations may be matched either
by: high extinction ($A_V=10$) clouds with moderate densities and FUV fields
($G_0\sim 10^3$) but with  $n_{\rm CO}>n_{\rm CII}$
so that the area of CO emitting gas is effectively decreased relative to the 
area of [C~II] emitting gas; or by lower extinction ($A_V\sim 3$) gas 
illuminated with $G_0\sim 300$ with equal densities 
($n\sim 10^4\,\rm cm^{-3}$) in the CO and [C~II] emitting regions, 
which again tends to lower the ratio of CO-to-[C~II] emitting areas. 

If the 
CO and [C~II] emitting regions have the same density, then our results from 
Figure 17 allow us to constrain the implied value of the ratio $L[{\rm CO}\,
J=1-0]$/M(H$_2$). For low metallicity, $n\sim 10^4\,\rm cm^{-3}$, and $G_0\sim
300$, and taking $A_V\sim 3$, Figure 20 shows that the ratio has a value of 
$\sim 3\times 10^{-9}$, a factor of $\sim 2\times 10^3$ smaller than the 
standard ratio. This low ratio implies that the mass of clouds in the LMC 
is larger by this factor than the standard ratio would give. Using this 
conversion factor, we can calculate the mass of H$_2$ in the 1\arcmin~ beam
from the observed CO intensity. If we assume that typical GMCs are $\sim$15 pc
in size, then the 1\arcmin~ beam contains a single GMC. We find a mass
M(H$_2) \sim 2.5\times 10^7\,\rm M_{\odot}$, more than an order of 
magnitude too high for a 15 pc cloud to be gravitationally stable. This seems
to rule out our $A_V=3$ model, so we conclude that the $A_V=10$ model with 
a density contrast between the CO and [C~II] emitting regions is 
far more likely. The $A_V=10$ model is consistent with the McKee (1989) 
model of photo-ionization regulated star formation which predicts that 
$A_V$ remains constant even as $Z$ changes.   
  
\section{Summary}

We have modeled the emission from photodissociation regions using the
PDR code of TH85, but updated through the use of the most up to date values of
atomic and molecular data, chemical rate coefficients, and grain photoelectric
heating rates. We find that the new grain heating terms produce the largest
differences between our results and earlier results of our group. We have
presented contour plots showing the line intensity and/or intensity ratios
for the PDR diagnostic lines [C~II] 158$\mu$m, [O~I] 63$\mu$m and 145$\mu$m, 
[C~I] 370$\mu$m and 609$\mu$m, CO $J=1-0$, $J=2-1$, $J=3-2$, $J=6-5$ and 
$J=15-14$, 
as well as the strength of the FIR continuum. The contour plots
cover a large range of density and incident FUV flux, including conditions
relevant to diffuse galactic PDR emission, the surfaces of GMCs exposed to the
average interstellar field, planetary nebulae, reflection nebulae and PDRs 
lying outside dense HII regions around O and B stars. These plots can be used
to derive the gas density $n$, incident FUV flux $G_0$, and gas temperature 
$T$
in PDR regions from observations of line fluxes.  

We have explored the effects of metallicity, cloud geometry and visual 
extinction on the [C~II]/FIR and CO $J=1-0$/FIR intensity ratios. 
We find that reducing the metallicity in a cloud
of fixed $A_V$ ($\sim$ 10) has little effect on the [C~II]/CO line ratio
except in the parameter regime $G_0/n < 3\times 10^{-3}\,\rm cm^3$, where the
[C~II]/CO ratio increases with decreasing $Z$ (see Fig. 16).
For translucent clouds ($A_V \sim 1-3$) changes in metallicity have a more
pronounced effect than in the $A_V\sim 10$ case (see Fig. 17 and 18). We find
 that
reducing the $A_V$ of a cloud greatly increases the  observed [C~II]/CO line 
ratio. Therefore, reducing the metallicity in a cloud of given total hydrogen
column density increases this ratio.
Increasing the density of the CO--emitting gas relative to that of the 
[C~II]--emitting gas in an unresolved spherical cloud produces an increase in
this ratio as well, because the area of the CO photosphere (and consequently
the CO luminosity) is reduced.  

We have applied our results to recent observations of the central $\sim$500 
pc 
of the starburst galaxy M82. 
Using the cloud ensemble technique from WTH90 together with our new model 
results, we find that the core of M82 contains clouds with
 the following average properties:
gas density, $n\sim 10^4\,\rm cm^{-3}$; FUV flux, $G_0=10^{3.5}$;
surface temperature, $T_S=400$~K; total mass of atomic gas in all the cloud
surfaces, $M_a\sim 8.8\times 10^6
\,M_{\odot}$; number of molecular clouds, $N_{cl}\sim 7500$; cloud mass,
$M_{cl}\sim 3\times 10^4\,M_{\odot}$; and cloud radii, $r_{cl}\sim 1.7$~pc. 
These clouds have masses akin to GMCs  in the Galaxy, though they
are denser and at pressures of order 100--1000 times that of local GMCs.  
Our derived values for the gas densities
of these clouds and the mean incident FUV flux are similar to prevously
determined values. On the other hand,  for emission over $\sim$4 kpc of the
spiral galaxy NGC 278, we find typical  values of: $n\sim 2\times 10^3\,\rm
cm^{-3}$; $G_0=70$; $T_S=200$~K; $M_a\sim 5\times 10^7
\,M_{\odot}$; $N_{cl}\sim 3\times 10^4$; $M_{cl}\sim 3.5\times
10^4\,M_{\odot}$; and $r_{cl}\sim 5.5$~pc. The lower values of $n$ and $G_0$
are expected since we are averaging over a greater volume of the galaxy and
because NGC 278 is not a starburst galaxy like M82. We find that the recently
observed high  [C~II]/CO ratio in the 30 Doradus region of the LMC can be
explained by either a lower $A_V$ ($A_V\sim 3$) in the typical clouds compared
with clouds in the Milky Way Galaxy ($A_V\sim 10$), or by a higher density in
the CO emitting  cloud cores compared with the [C~II] halos. In either case,
the column of  hydrogen nuclei through the typical clouds is
greater than the  columns through Milky Way molecular clouds. If the [C~II]/CO
ratio in 30  Doradus is caused by $A_V\sim 3$ clouds, then the CO-to-H2
conversion factor for these clouds is reduced by a factor of as much as $\sim
10^3$  compared to Milky Way clouds. This ratio gives an H$_2$ mass for GMCs 
in the LMC an order of magnitude too 
high for the clouds to be gravitationally stable, and so we rule out
low $A_V$ clouds as an explanation for the high [C~II]/CO ratio. 
Instead, we find it more likely that these GMCs have $A_V=10$, 
consistent with the McKee (1989) model of photo-ionization regulated 
star formation which predicts that $A_V$ remains constant even as $Z$ 
changes.

\acknowledgements
We would like to thank the referee, Amiel Sternberg, for making 
helpful suggestions which impoved the manuscript. M.~J.~K. and D.~J.~H. 
acknowledge NASA RTOP 
399-28-01-07, which supports the ISO Key Project on Normal Galaxies,  
and NASA RTOP 399-25-01, which supports SWAS. M.~G.~W. was supported in part by
a NASA-Ames University Consortium grant NCC-25237 and by NASA ADP grant 
NAG-56750. 
\clearpage

\begin{table}
\caption{Standard Model Parameters}
\smallskip\smallskip
\begin{center}
\begin{tabular}{lll}
\tableline
\tableline
Parameter&Symbol&Value\\
\tableline
Turbulent Doppler Velocity&$\delta v_D$&1.5$\,\rm km\,s^{-1}$\\
Carbon abundance\tablenotemark{a}&$x_C$&$1.4\times 10^{-4}$\\
Oxygen abundance\tablenotemark{a}&$x_O$&$3.0\times 10^{-4}$\\
Silicon abundance\tablenotemark{a}&$x_{Si}$&$1.7\times 10^{-6}$\\
Sulfer abundance\tablenotemark{a}&$x_{S}$&$2.8\times 10^{-5}$\\
Iron abundance\tablenotemark{a}&$x_{Fe}$&$1.7\times 10^{-7}$\\
Magnesium abundance\tablenotemark{a}&$x_{Mg}$&$1.1\times 10^{-6}$\\
Dust abundance relative to diffuse ISM&$\delta_d$&1\\
FUV Dust absorption/visual extinction&$\delta_{UV}$&1.8\\
Dust visual cross section per H&$\sigma_{\rm V}$&$5\times 10^{-22}\,\rm cm^{2}$\\
Formation rate of $\rm H_2$ on dust\tablenotemark{b}&$R_{form}$&$3\times 10^{-17}\,\rm cm^3\,s^{-1}$\\
PAH abundance&$x_{\rm PAH}=n_{\rm PAH}/n$&$4\times 10^{-7}$\\
Cloud density&$n$&$10^1\,{\rm cm^{-3}}\le\,n\,\le\, 10^7\,{\rm cm^{-3}}$\\
Incident FUV flux&$G_0$&$10^{-0.5}\,\le\,G_0\,\le\, 10^{6.5}$\\
\tableline
\end{tabular}
\end{center}
\tablenotetext{a} {Gas phase abundances relative to H nuclei from 
Savage \& Sembach (1996)}
\tablenotetext{b} {Rate per unit volume (cm$^{-3}$ s$^{-1}$) given by $R_{form}n_{\rm HI}n$}
\end{table}

\clearpage

\begin{table}
\caption{PDR Diagnostic Transitions}
\begin{center}
\begin{tabular}{llrrr}
\tableline
\tableline
Species&Transition&Wavelength&$E_{upper}/k$\hfill&$n_{cr}\tablenotemark{a}$\hfill\\
&&[$\mu$m]&[$K$]&[cm$^{-3}$]\\
\tableline
$[$C~II$]$&$^2P_{3/2}-^2P_{1/2}$&157.74&92&$3\times 10^3$\\
$[$O~I$]$&$^3P_{1}-^3P_{2}$&63.18&228&$5\times 10^5$\\
$[$O~I$]$&$^3P_0-^3P_1$&145.53&326&$1\times10^5$\\
$[$C~I$]$&$^3P_1-^3P_0$&609.14&23.6&$5\times10^2$\\
$[$C~I$]$&$^3P_2-^3P_1$&369.87&62.5&$3\times 10^3$\\
CO&$J=1-0$&2600.78&5.53&$3\times10^3$\\
CO&$J=2-1$&1300.39&16.59&$1\times10^4$\\
CO&$J=3-2$&867.00&33.19&$5\times10^4$\\
CO&$J=6-5$&433.56&116.16&$4\times10^5$\\
CO&$J=15-14$&173.62&663.36&$8\times10^6$\\
\tableline
\end{tabular}
\end{center}
\tablenotetext{a} {Critical density for [C~II], [O~I], and [C~I] for 
collisions with H. Critical density for CO for collisions with H$_2$.
Critical densities for atoms ([C~II], [O~I] \& [C~I]) from
TH85. Critical densities for CO calculated from $n_{cr}=A_{ul}/\gamma_{ul}$
for a temperature of 100~K, $A$--values from Rothman \ea (1987)
and values of $\gamma_{ul}$ from Viscuso \& Chernoff (1988).} 
\end{table}

\clearpage

\begin{figure}
\figurenum{1}
\plotone{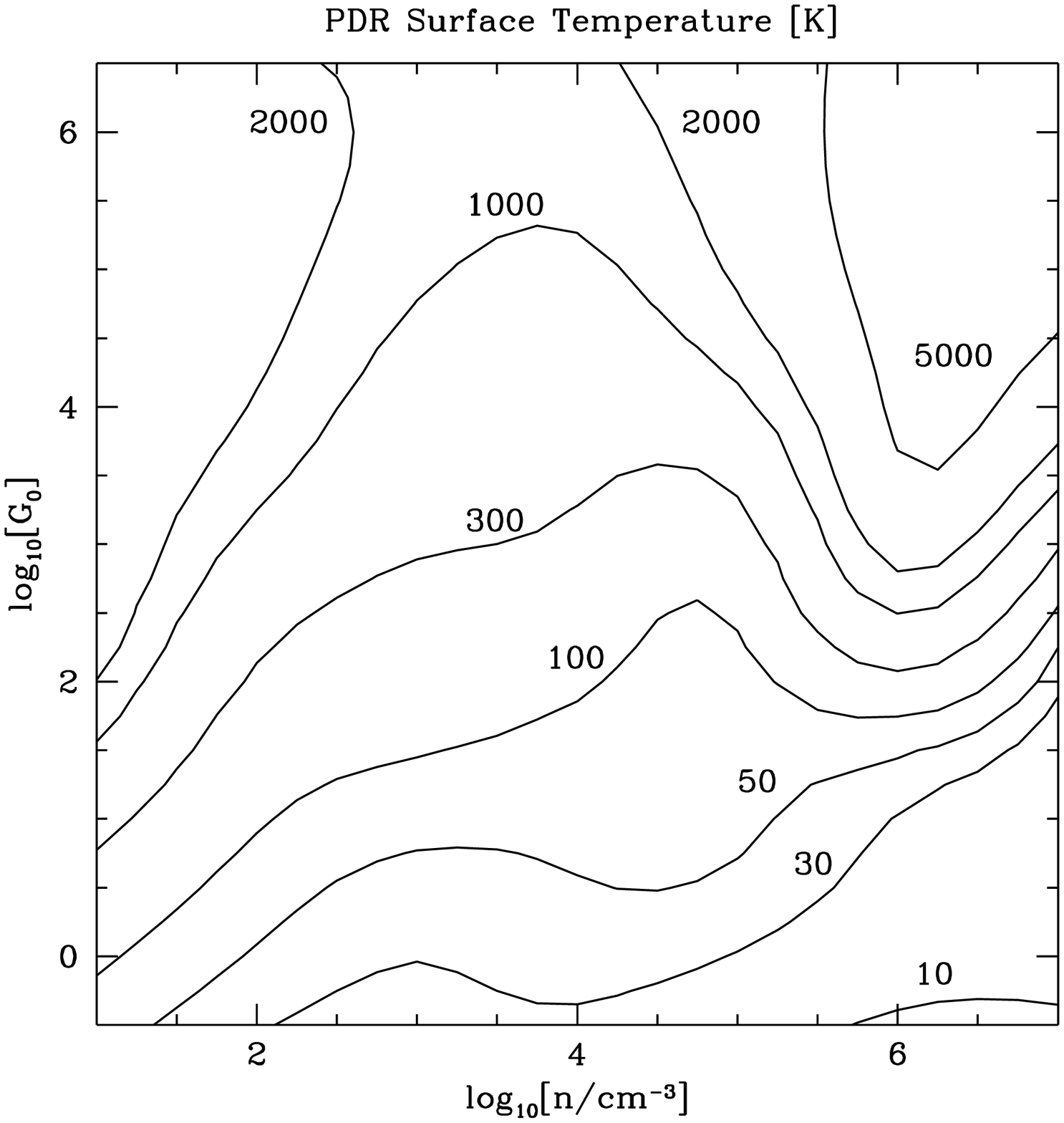}
\caption{Surface temperature of the atomic gas as a function of $n$ and $G_0$.
Contour levels are in units of degrees Kelvin.}
\end{figure}

\begin{figure}
\figurenum{2}
\plotone{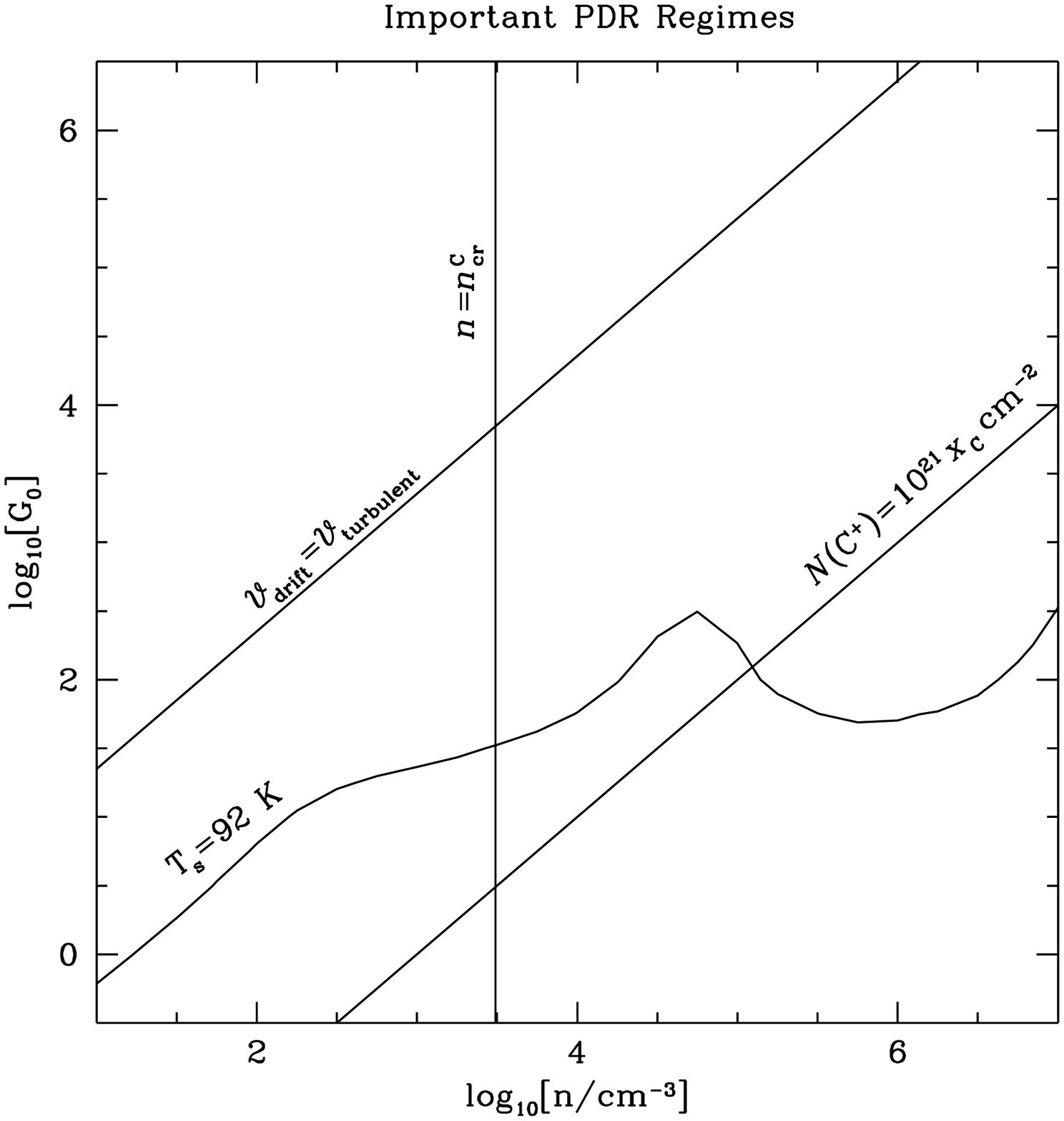}
\end{figure}
\clearpage
\begin{figure}
\figurenum{2}
\caption{Important regimes for determining the intensity of a cooling line 
in a PDR. The line $n=n_{cr}^{\rm C}$ roughly marks the critical density for the [C~II] 158
$\mu$m, CO $J=1-0$, and [C~I] 370$\mu$m transitions. For optically thin lines 
at constant column density, line intensities increase as $\propto\,n^1$ for 
$n<n_{cr}$ and as $\propto\,n^0$ for $n>n_{cr}$. The line $v_{\rm drift}=v_{\rm turbulent}$
shows where radiation pressure causes grains to drift with a speed equal to the 
gas turbulent speed. Parameter space above this line is inconsistent with
a steady state model. The curve $T_S=92$~K shows where the surface gas temperature 
is equal to the [C~II] energy level above ground ($kT_S=\Delta E$). Above this curve
the [C~II] intensity, for constant column density, is weakly dependent on $G_0$. The 
line $N({\rm C^+})=10^{21}x_{\rm C}$ traces the location of a constant \cp~ column 
density ($\sim 10^{17}\,\rm cm^{-2}$), equivalent to all carbon being singly ionized to a 
hydrogen column density of $10^{21}\,\rm cm^{-2}$. Below this line
gas shielding of the FUV field becomes important and the \cp~ column density and the [C~II] line intensity
become sensitive to $G_0/n$.} 
\end{figure}
\clearpage

\begin{figure}
\figurenum{3}
\plotone{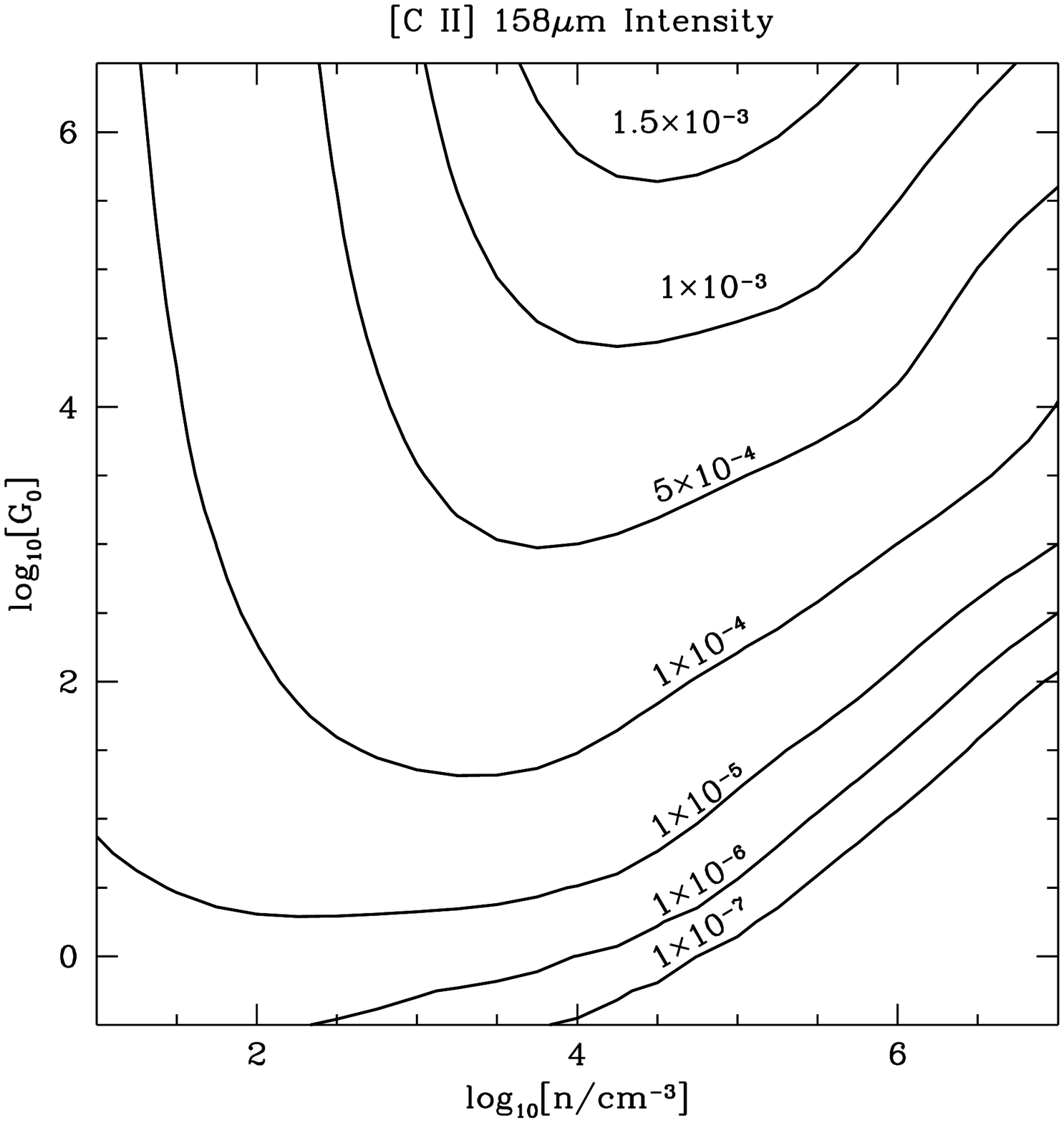}
\caption{[C~II] 158$\mu$m intensity emitted from the surface of a photodissociation 
region as a function of the cloud density, $n$, and the FUV flux incident on 
the cloud, $G_0$, for our standard model parameters. Contours are labeled with
the line intensity in units of erg cm$^{-2}$ s$^{-1}$ sr$^{-1}.$}
\end{figure}

\begin{figure}
\figurenum{4}
\plotone{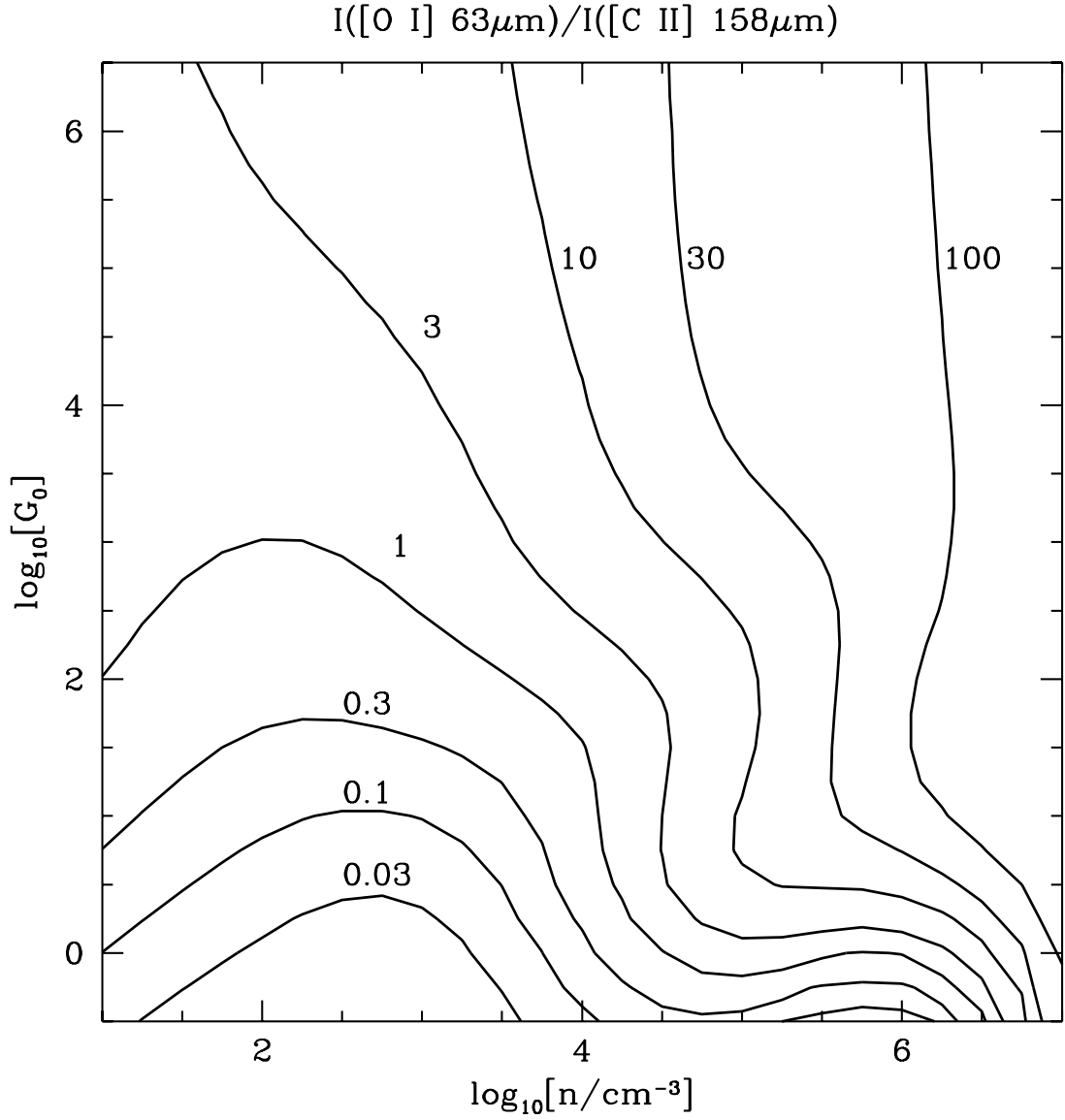}
\caption{Ratio of the intensity of the [O~I] 63$\mu$m line to the intensity of the
[C~II] 158$\mu$m line
emitted from the surface of a photodissociation 
region as a function of the cloud density, $n$, and the FUV flux incident on 
the cloud, $G_0$, for our standard model parameters.}
\end{figure}

\begin{figure}
\figurenum{5}
\plotone{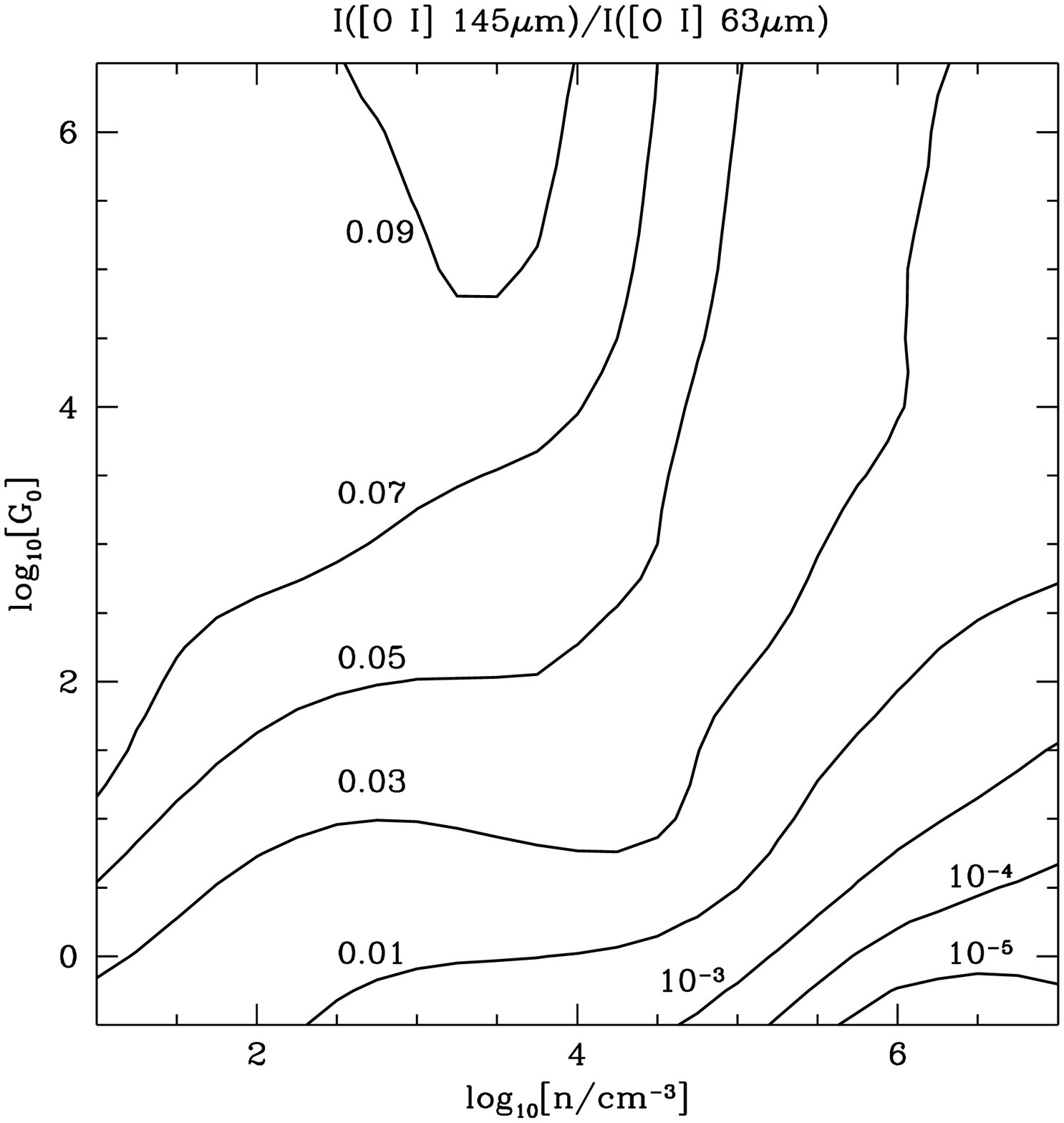}
\caption{Ratio of the intensity of the [O~I] 63$\mu$m line to the [O~I] 145$\mu$m line
as a function of the cloud density, $n$, and the FUV flux incident on 
the cloud, $G_0$, for our standard model parameters.}
\end{figure}

\begin{figure}
\figurenum{6}
\plotone{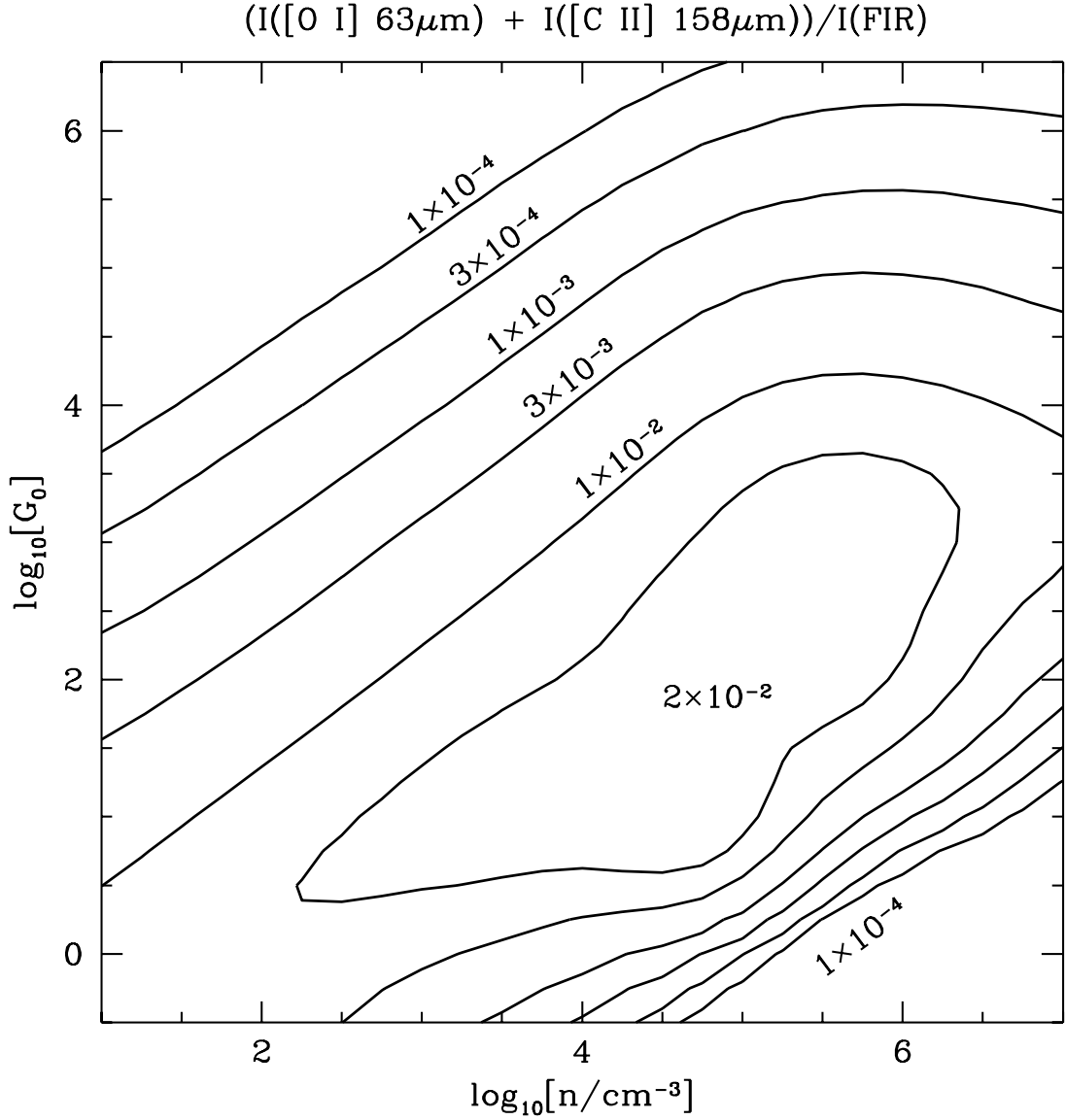}
\caption{Ratio of the intensity of the [C~II] 158$\mu$m and [O~I] 63$\mu$m lines to the
total far infrared intensity emitted from the surface of a photodissociation
region as a function of the cloud density, $n$, and the FUV flux incident on 
the cloud, $G_0$, for our standard model parameters. See text for cautions on using this figure.}
\end{figure}

\begin{figure}
\figurenum{7}
\plotone{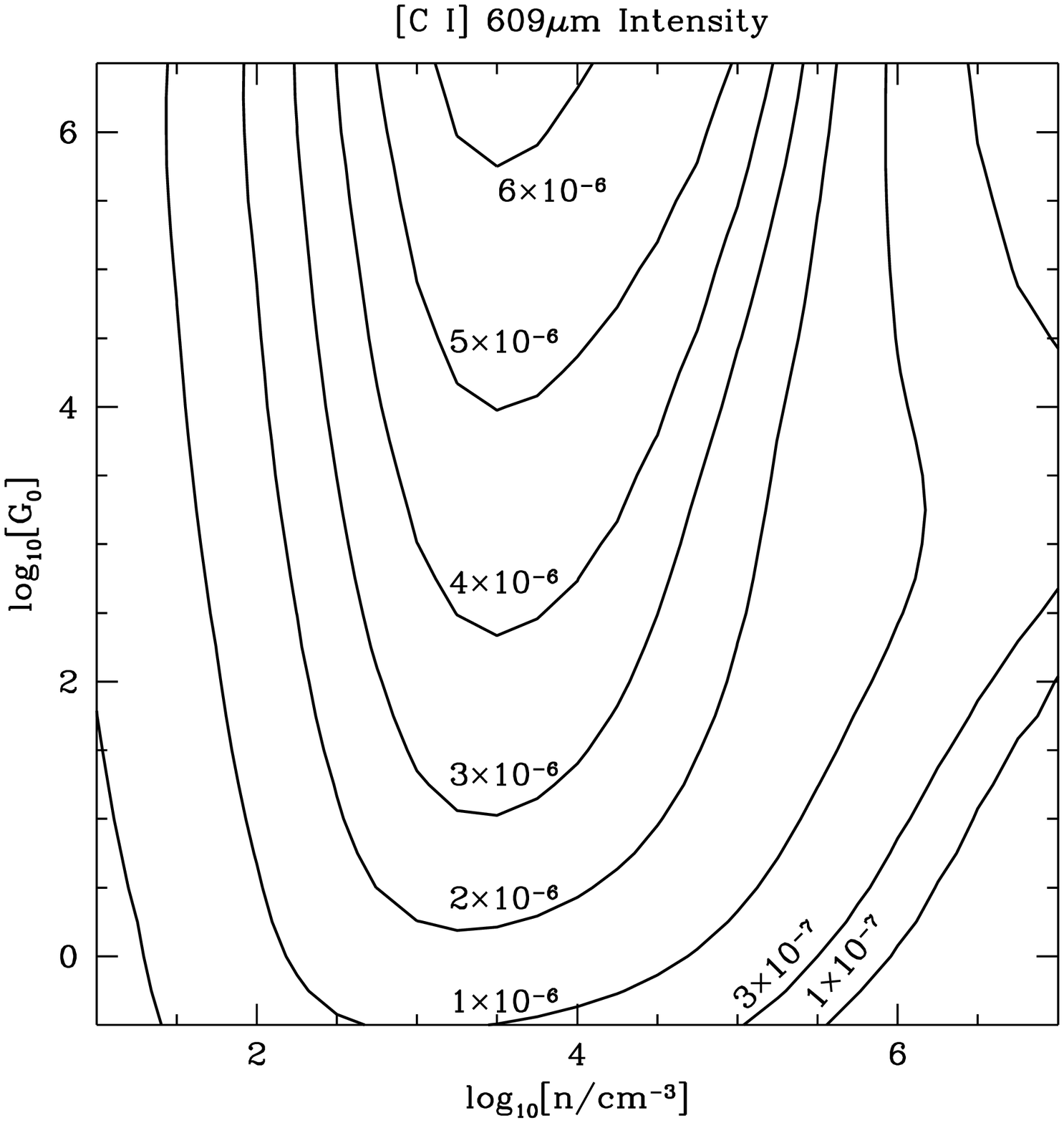}
\caption{[C~I] 609$\mu$m intensity emitted from the surface of a photodissociation 
region as a function of the cloud density, $n$, and the FUV flux incident on 
the cloud, $G_0$, for our standard model parameters. Contours are labeled with
the line intensity in units of erg cm$^{-2}$ s$^{-1}$ sr$^{-1}.$}
\end{figure}

\begin{figure}
\figurenum{8}
\plotone{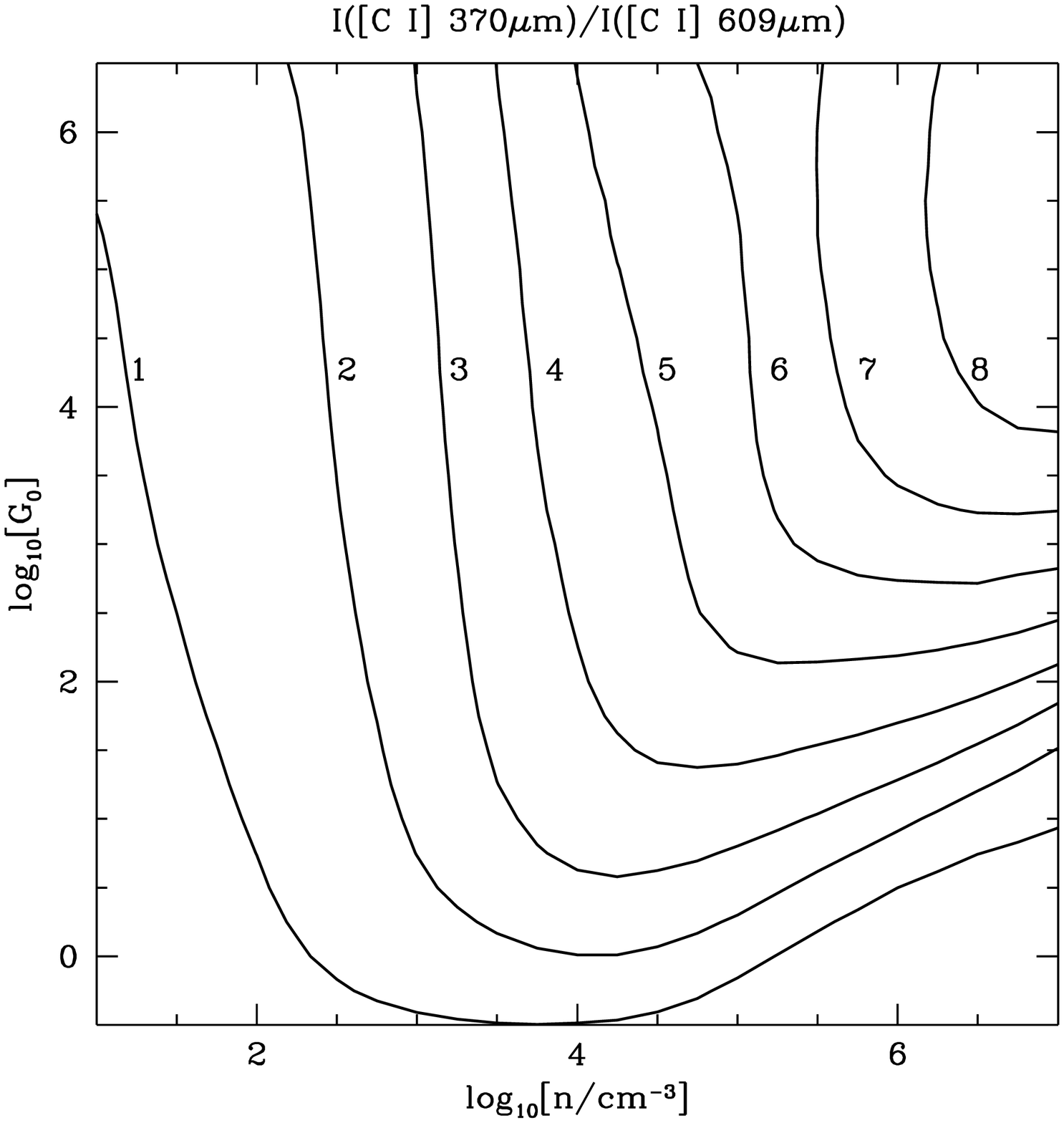}
\caption{Ratio of the intensity of the [C~I] 370$\mu$m line to the [C~I] 609$\mu$m line
as a function of the cloud density, $n$, and the FUV flux incident on 
the cloud, $G_0$, for our standard model parameters.}
\end{figure}
\clearpage

\begin{figure}
\figurenum{9}
\plotone{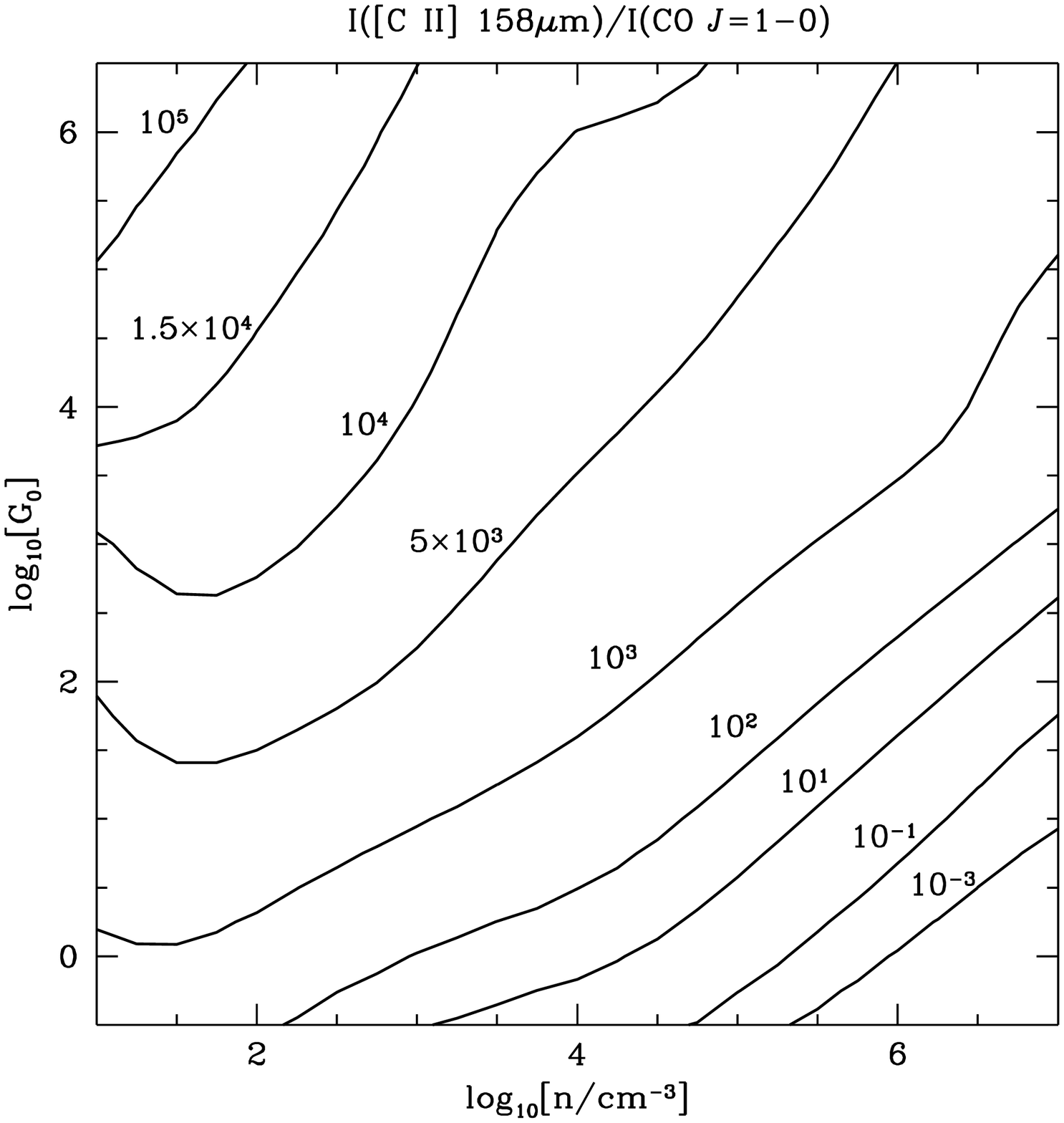}
\caption{Ratio of the intensity of the [C~II] 158$\mu$m line to the CO $J=1-0$ 2.6mm
line as a function of the cloud density, $n$, and the FUV flux incident on 
the cloud, $G_0$, for our standard model parameters, with assumptions noted in \S3.3.}
\end{figure}

\begin{figure}
\figurenum{10}
\plotone{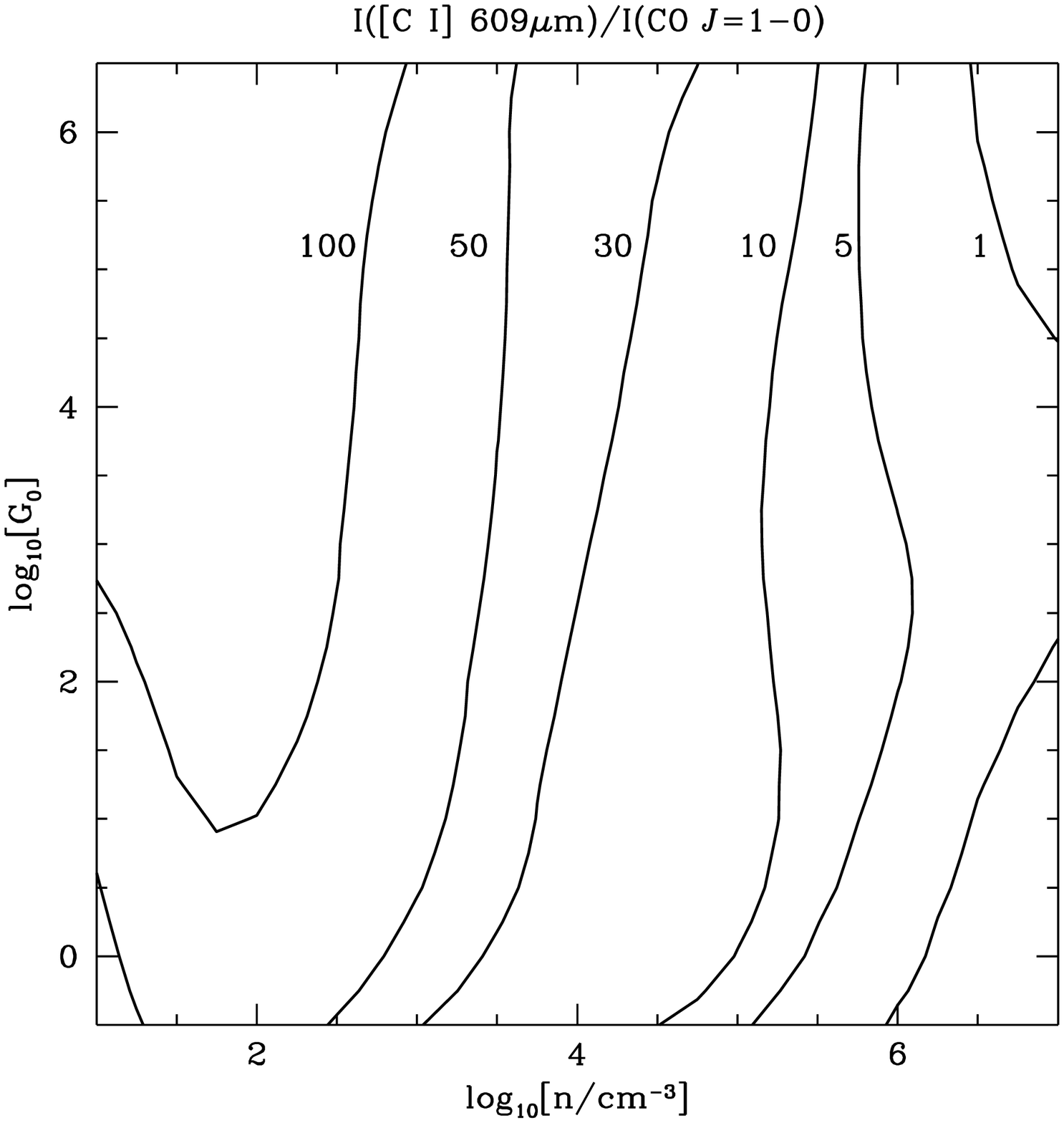}
\caption{Ratio of the intensity of the [C~I] 609$\mu$m line to the CO $J=1-0$ 2.6mm
line as a function of the cloud density, $n$, and the FUV flux incident on 
the cloud, $G_0$, for our standard model parameters, with assumptions noted in \S3.3.}
\end{figure}

\begin{figure}
\figurenum{11}
\plotone{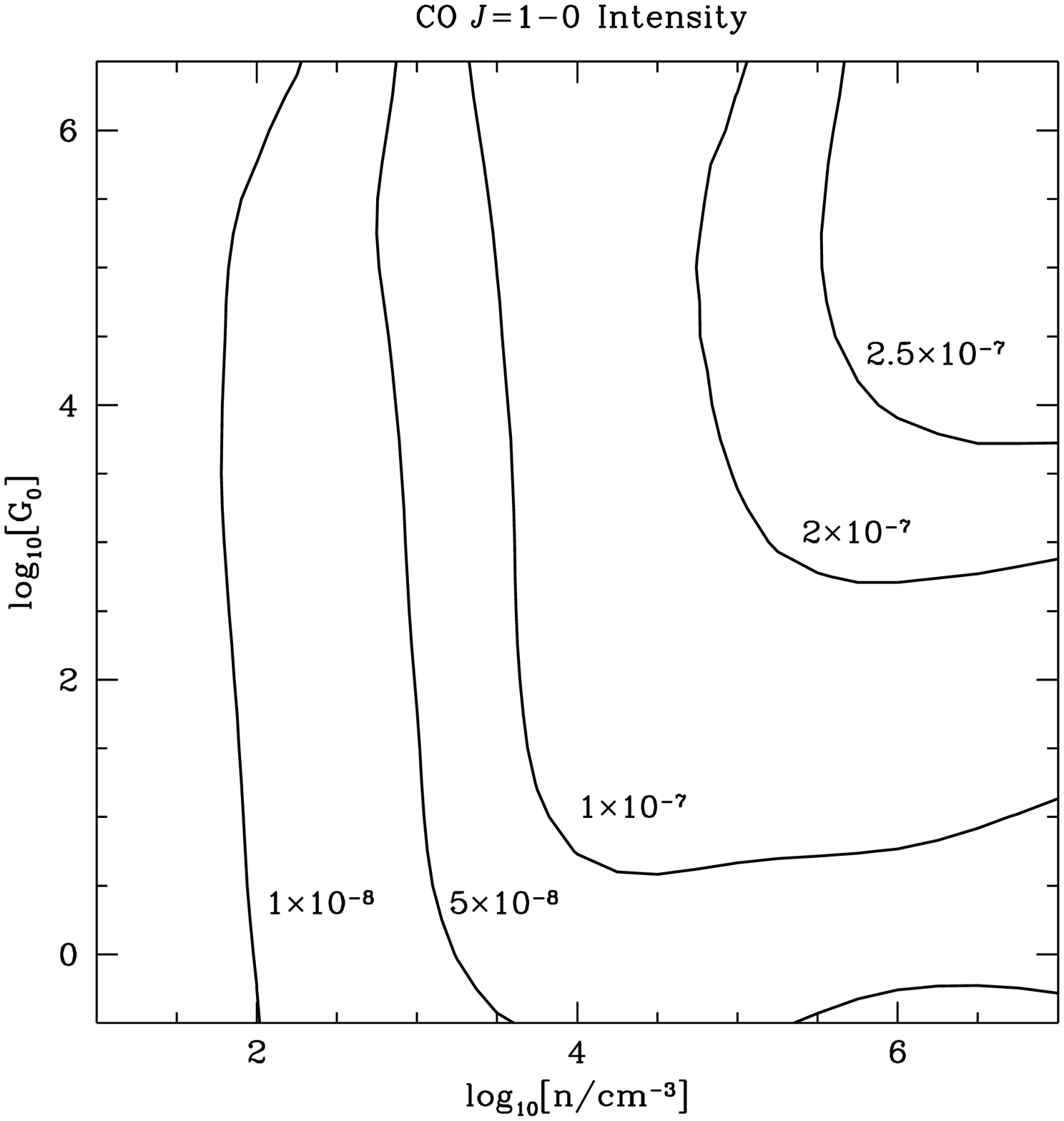}
\caption{CO $J=1-0$ 2.6mm intensity emitted from the surface of a photodissociation 
region as a function of the cloud density, $n$, and the FUV flux incident on 
the cloud, $G_0$, for our standard model parameters. Contours are labeled with
the line intensity in units of erg cm$^{-2}$ s$^{-1}$ sr$^{-1}.$}
\end{figure}

\begin{figure}
\figurenum{12}
\plotone{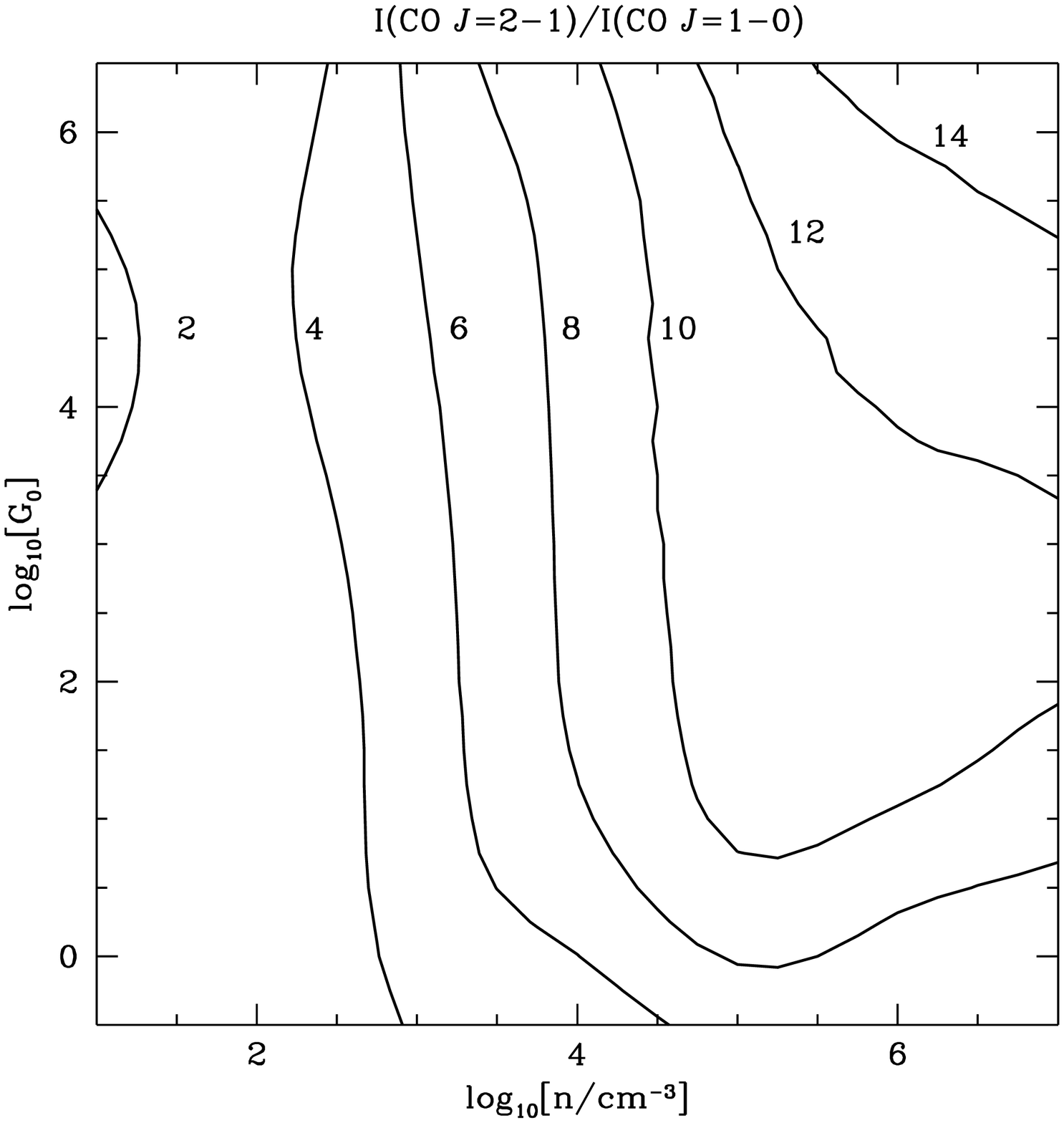}
\caption{Ratio of the intensity of the CO $J=2-1$ 1.3mm line to the CO 
$J=1-0$ 2.6mm line
as a function of the cloud density, $n$, and the FUV flux incident on 
the cloud, $G_0$, for our standard model parameters. Contours may be converted to antenna
temperature ratios by multiplying by 0.125, with assumptions noted in text.}
\end{figure}

\begin{figure}
\figurenum{13}
\plotone{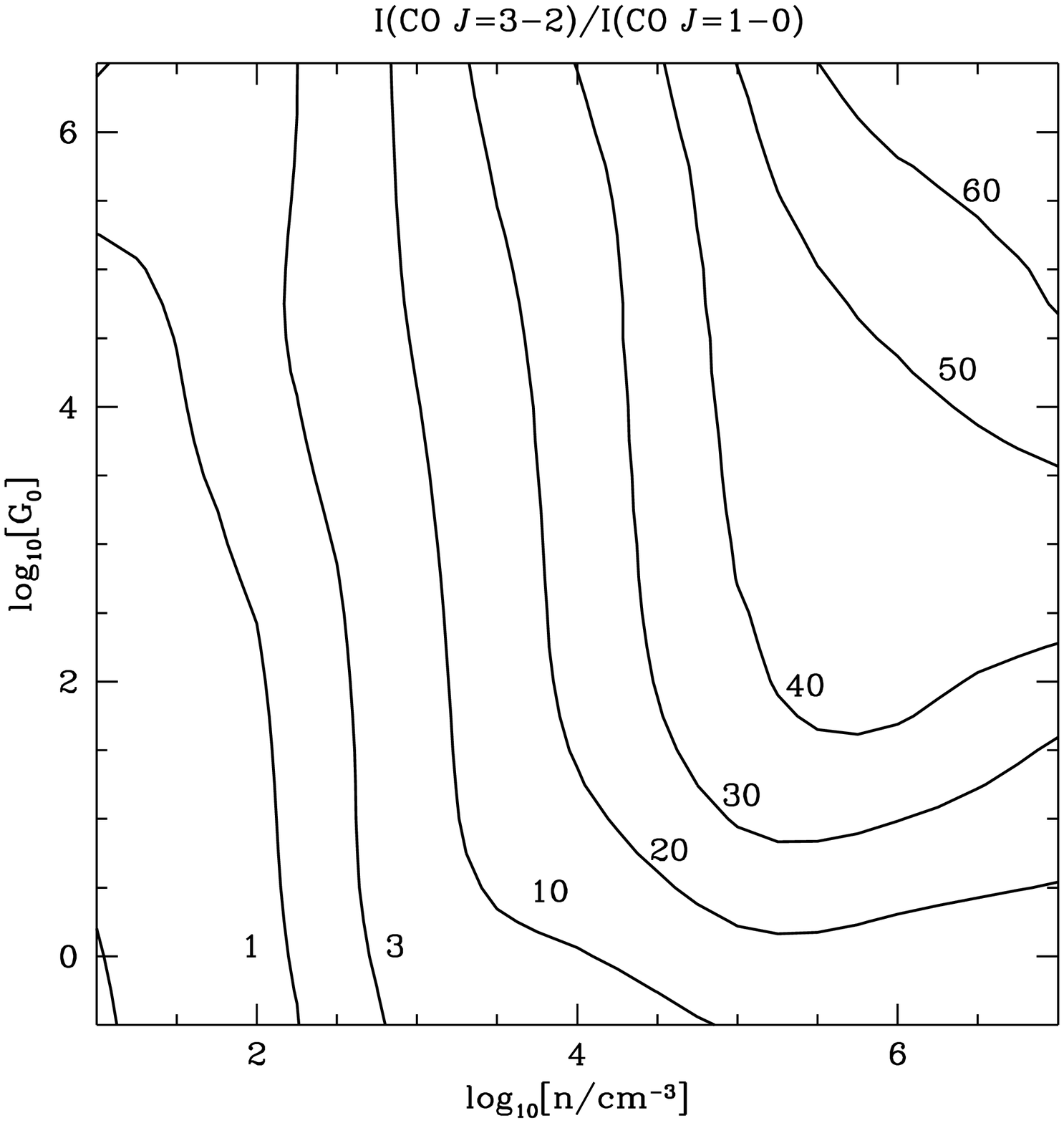}
\caption{Ratio of the intensity of the CO $J=3-2$ 867$\mu$m line to the CO 
$J=1-0$ 2.6mm line
as a function of the cloud density, $n$, and the FUV flux incident on 
the cloud, $G_0$, for our standard model parameters. Contours may be converted to antenna
temperature ratios by multiplying by 0.037, with assumptions noted in text.}
\end{figure}

\begin{figure}
\figurenum{14}
\plotone{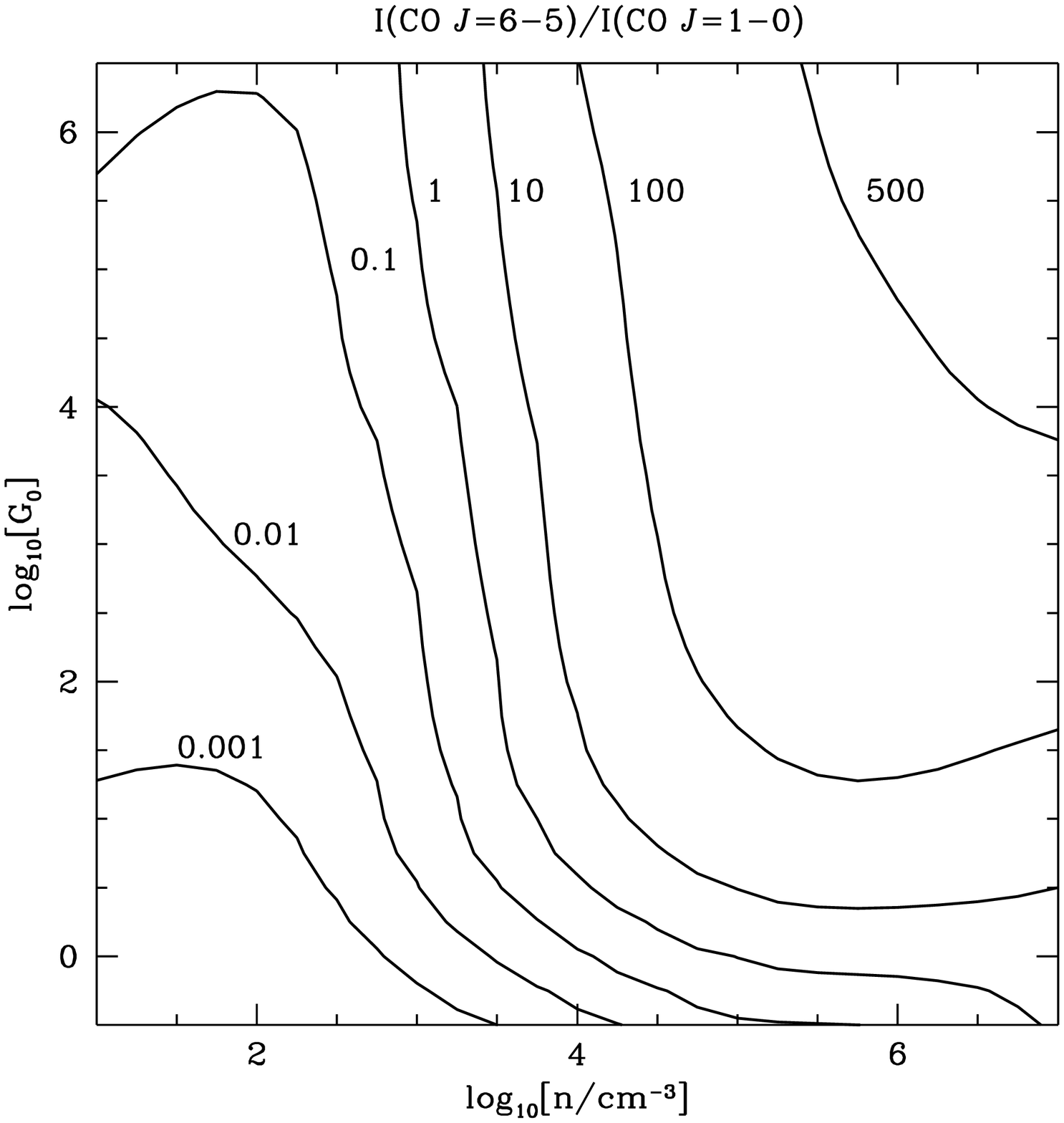}
\caption{Ratio of the intensity of the CO $J=6-5$ 433$\mu$m line to the CO 
$J=1-0$ 2.6mm line
as a function of the cloud density, $n$, and the FUV flux incident on 
the cloud, $G_0$, for our standard model parameters.}
\end{figure}

\begin{figure}
\figurenum{15}
\plotone{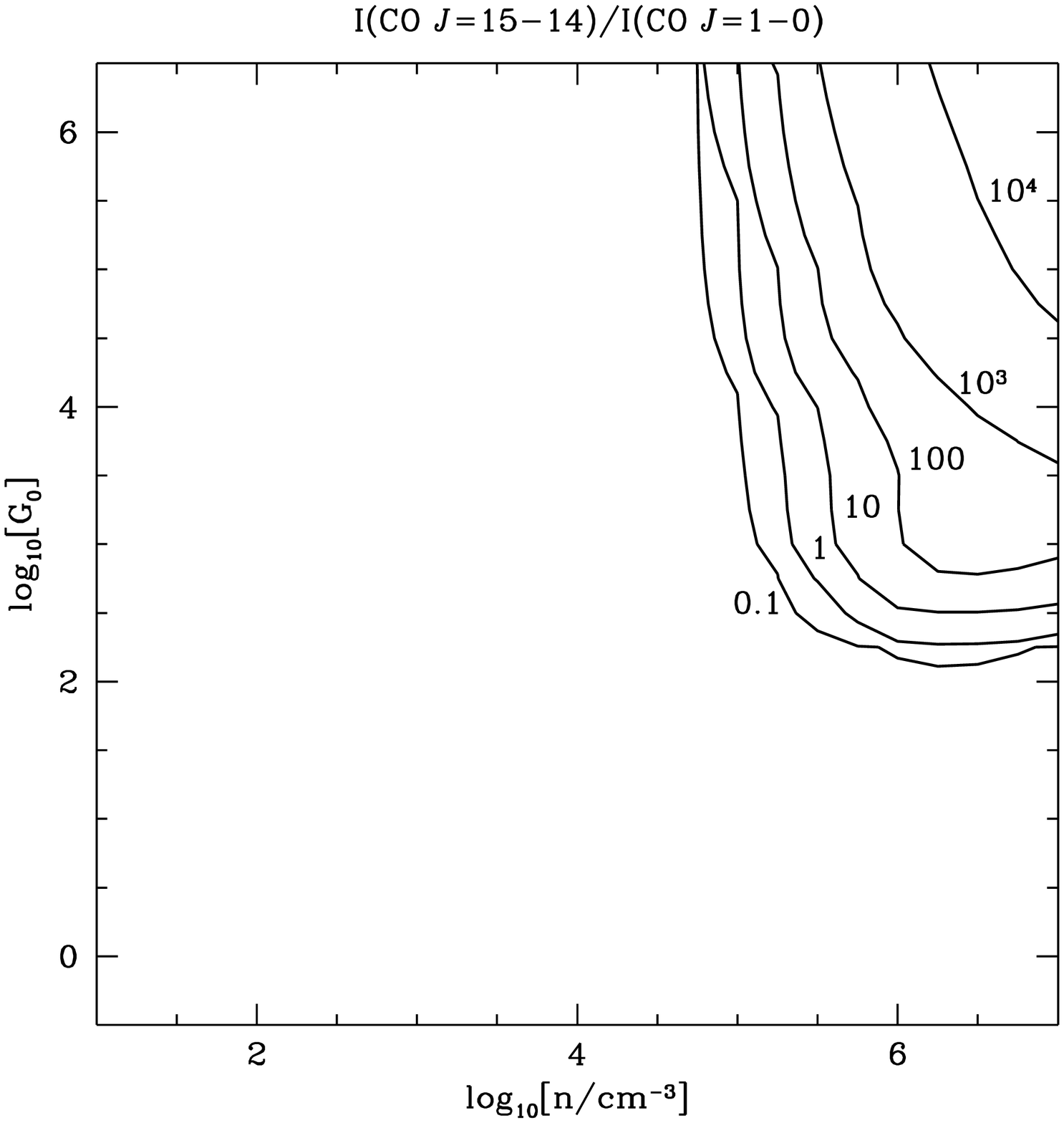}
\caption{Ratio of the intensity of the CO $J=15-14$ 173.6$\mu$m line to the CO 
$J=1-0$ 2.6mm line
as a function of the cloud density, $n$, and the FUV flux incident on 
the cloud, $G_0$, for our standard model parameters.}
\end{figure}
\clearpage

\begin{figure}
\figurenum{16}
\plotone{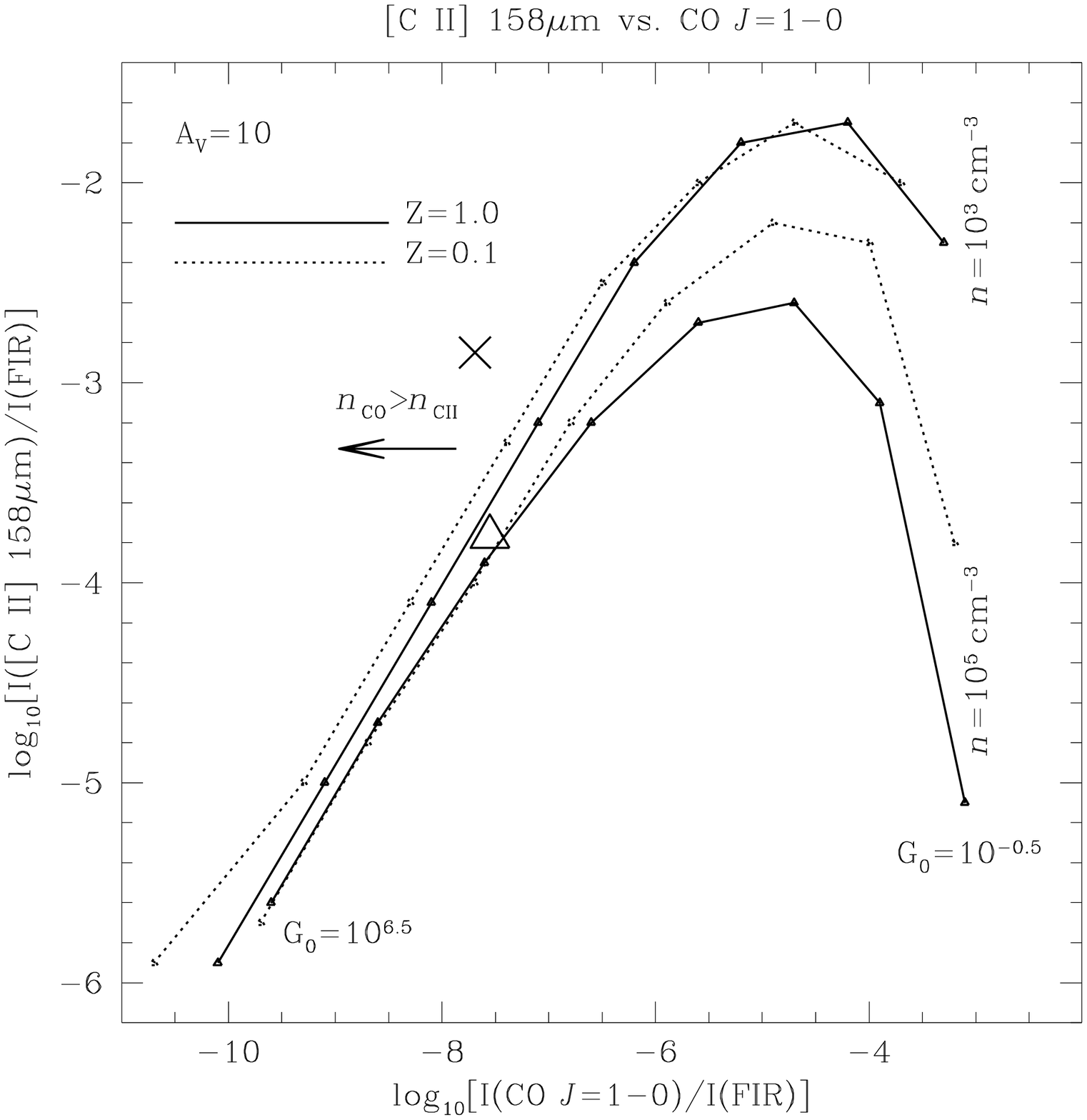}
\end{figure}
\clearpage

\begin{figure}
\figurenum{16}
\caption{Comparison of the [C~II] 158$\mu$m and CO $J=1-0$ transitions
scaled relative to the expected FIR intensity, for PDR models with a total
slab visual extinction $A_V=10$. Curves are labeled with density ($n=10^3\,
\rm cm^{-3}$ and $n=10^5\,\rm cm^{-3}$), and points
represent results for different values of the incident FUV radiation field.
The rightmost point on each curve is for log($G_0$)=--0.5, with $G_0$ 
increasing by a factor of 10 for each step to the left along a curve. The 
dotted lines $Z=0.1$ show
results for a factor of 10 reduction in the abundance of dust and metals. 
The arrow indicates the direction of the shift in curves with enhanced density in the CO zone relative to the [C~II] zone and for unresolved spherical clouds
instead of slab geometry. For 
$n_{\rm CO}\gg n_{\rm CII}$, the CO luminosity is small compared with $L({\rm C~II})$
even for $A_V=10$, because of the reduced area of the CO photosphere. The arrow
also indicates the shift which occurs for spherical clouds of smaller $A_V$ 
but with $n_{\rm CO}=n_{\rm CII}$. The $\rm X$ indicates the results for
the Large Magellenic Cloud (LMC), and the $\bf \triangle$ indicates the results for 
M82 (c.f. \S\S3.5.2-3.5.3).} 
\end{figure}
\clearpage

\begin{figure}
\figurenum{17}
\plotone{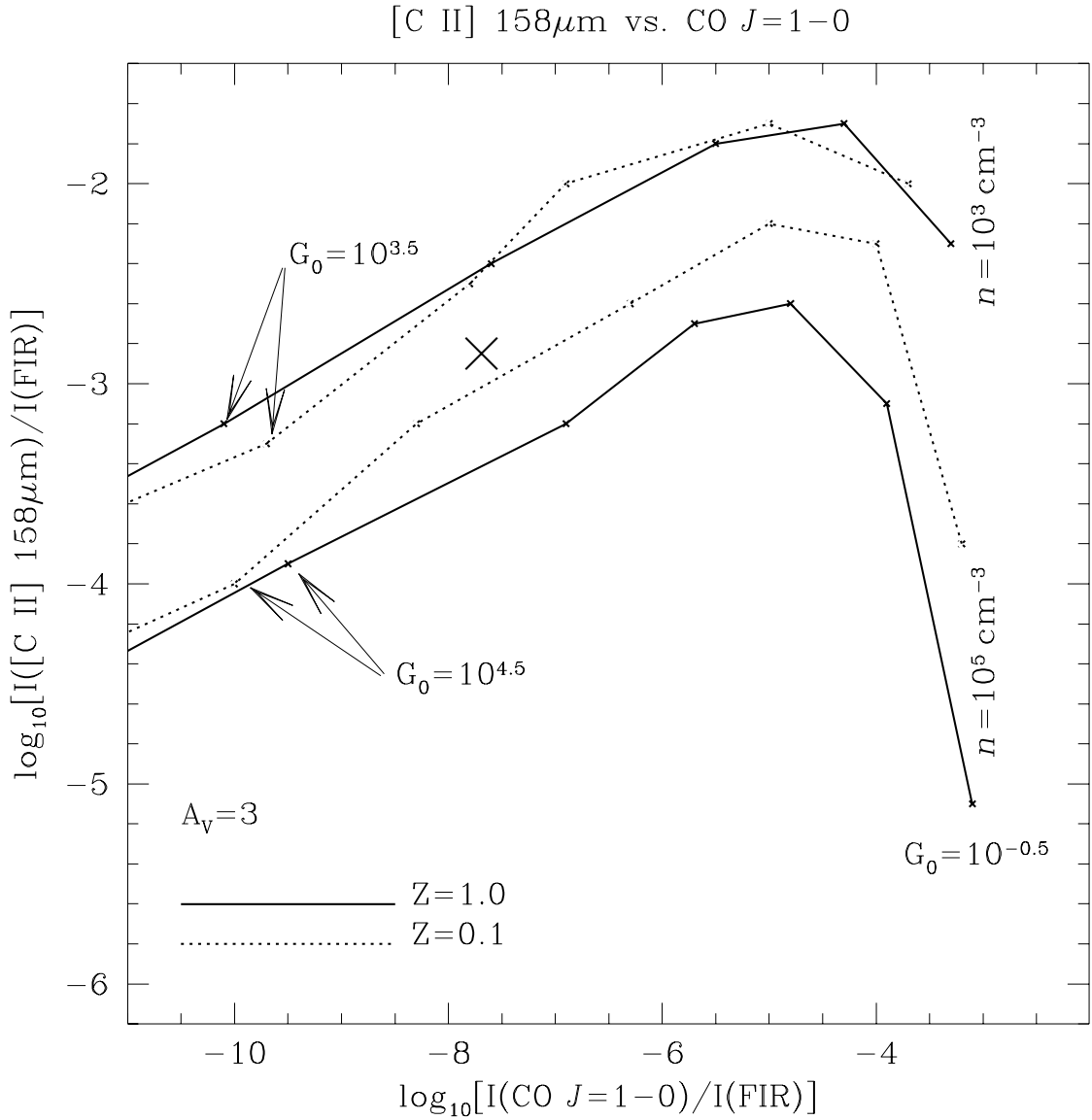}
\caption{Same as Figure 16, but for total slab visual extinction $A_V=3$.
Note that very high values of $G_0$ lie to the far left of the figure. Values
for the LMC are again indicated by the $\bf \times$.}
\end{figure}

\begin{figure}
\figurenum{18}
\plotone{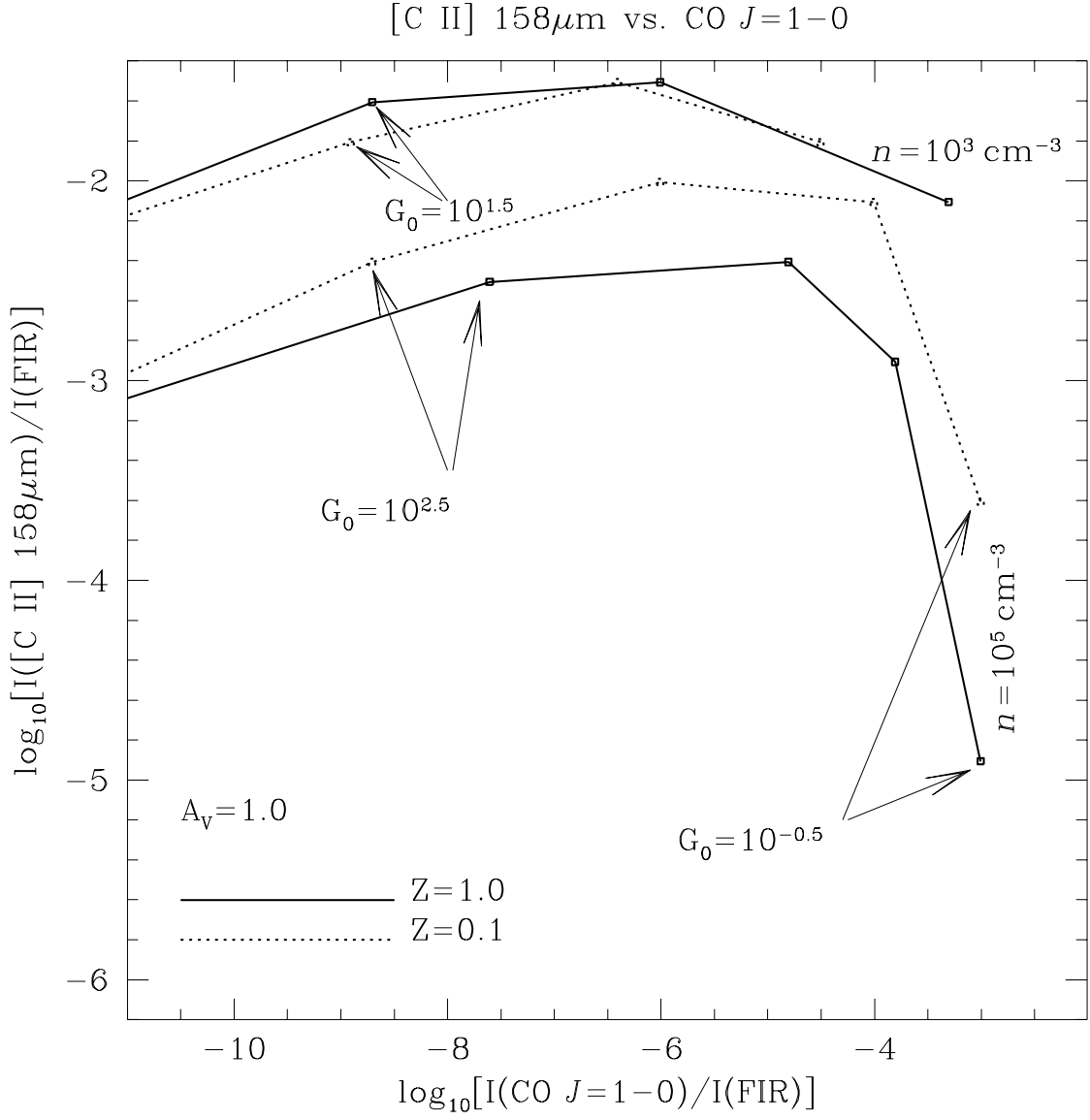}
\caption{Same as Figure 16, but for total slab visual extinction $A_V=1.0$.
Again, very high values of $G_0$ lie to the far left of the figure. The CO
intensity becomes very weak for high values of $G_0$ because most of the CO
is photodissociated and the slab becomes very optically thin in this line.}
\end{figure}

\begin{figure}
\figurenum{19}
\plotone{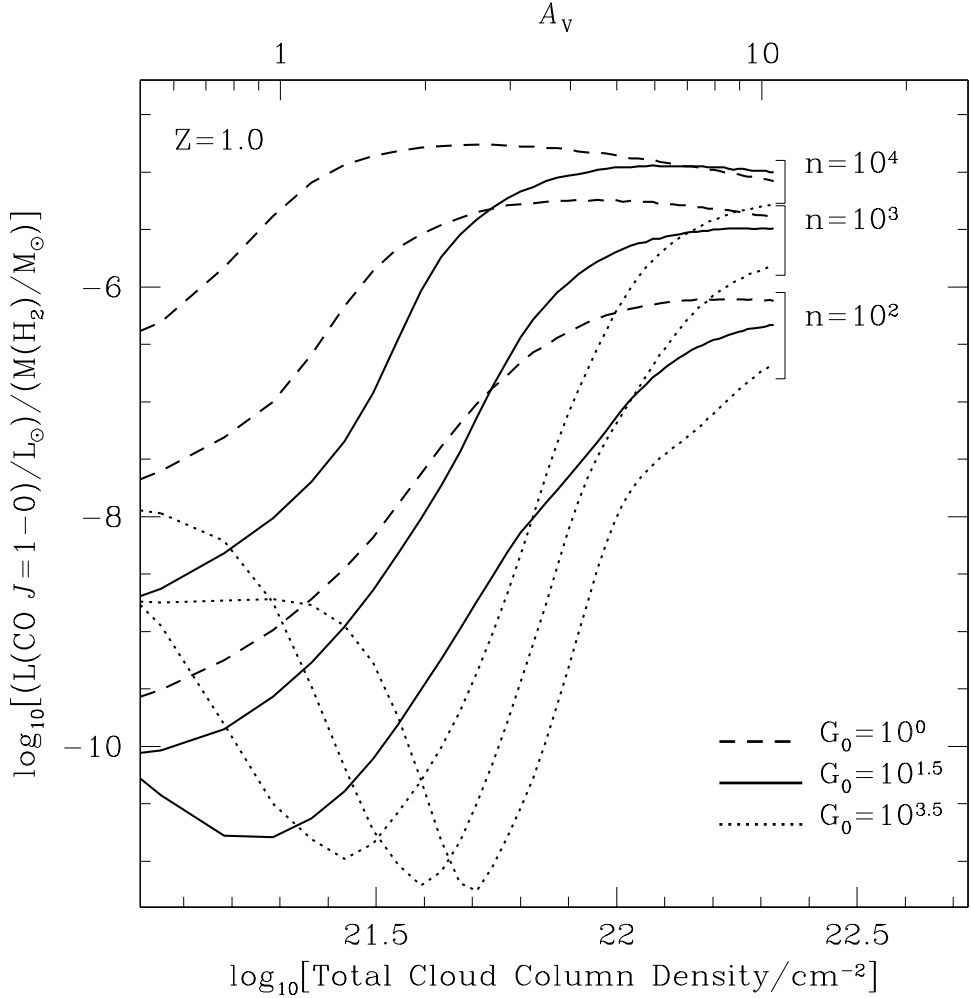}
\caption{CO $J=1-0$ luminosity (in units of $L_{\odot}$) divided by total 
H$_2$ mass (in units of M$_{\odot}$) for spherical PDR models with various
central optical depths or columns (see text). Results are presented for 
clouds with densities $n=10^2$, $10^3$ and $10^4\,\rm cm^{-3}$ and for FUV 
intensities $G_0=1$ (dashed curves), $10^{1.5}$ (solid curves) and $10^{3.5}$
(dotted curves). The standard conversion factor corresponds to a value $\sim
8\times 10^{-6}$ L$_{\odot}$/M$_{\odot}$, or $1.7\times 10^{-7}\, {\rm K\,
km\,s^{-1}\, kpc^2}$ per M$_{\odot}$.}
\end{figure}

\begin{figure}
\figurenum{20}
\plotone{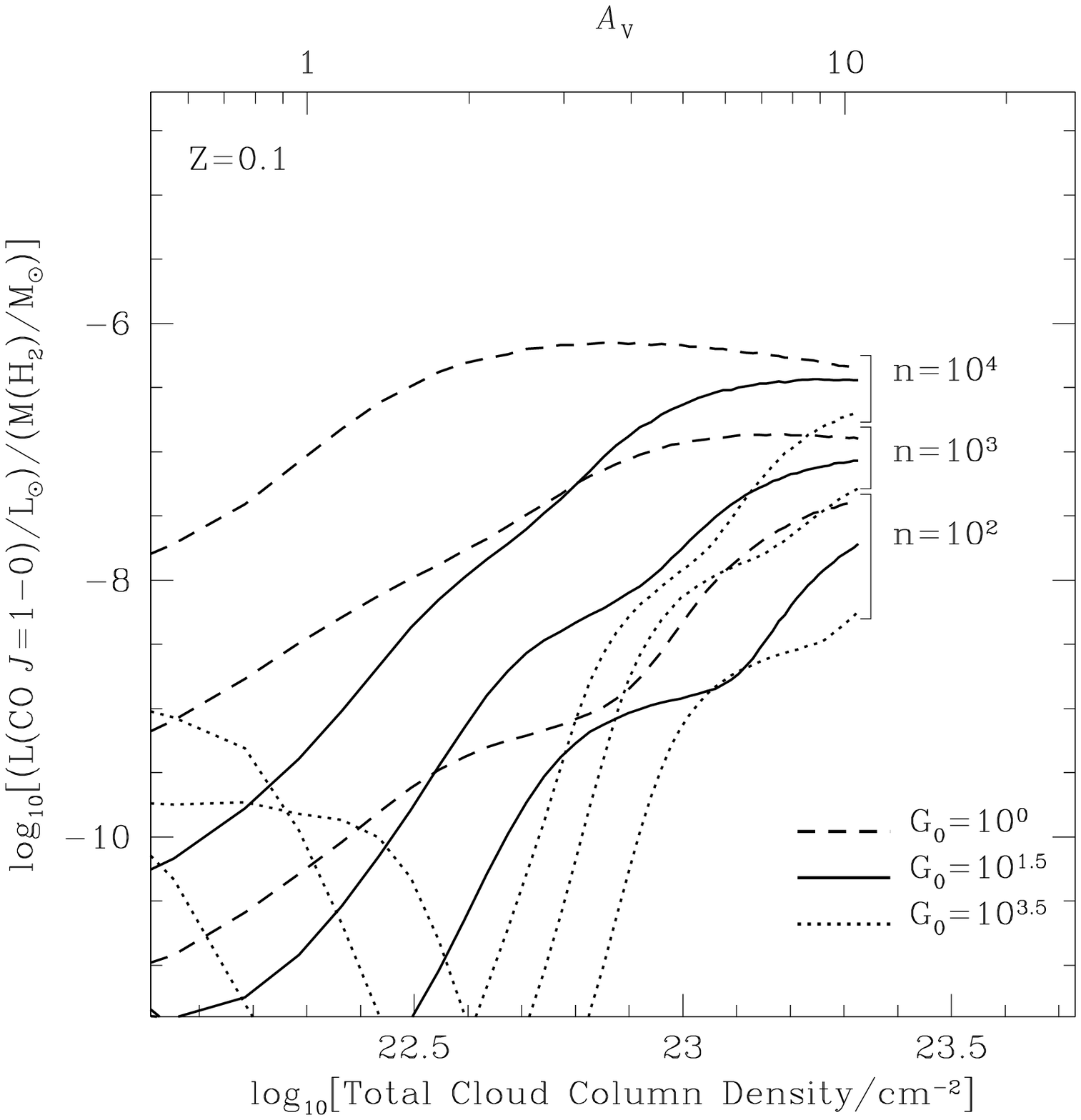}
\caption{Same as Figure 18, but for $Z=0.1$. Note the difference in the total 
cloud column density scale.}
\end{figure}

\clearpage


\clearpage
\begin{thebibliography}{}

\bibitem[]{} Abgrall, H., Le Bourlot, J., Pineau des For\^ets, G., Roueff, E., Flower, D. R., 
\& Heck, L. 1992, A\&A, 253, 525
\bibitem[]{}Bakes, E. L. O., \& Tielens, A. G. G. M. 1994, ApJ, 427, 822
\bibitem[]{}Bakes, E. L. O., \& Tielens, A. G. G. M. 1998, ApJ, 499, 258
%\bibitem[]{}Bennett, C.L., et al. 1994, ApJ, 434, 587
\bibitem[]{}Blum, R. D. \& Pradhan, A. K. 1992, ApJS, 80, 425
\bibitem[]{}Braine, J., Combes, F., Casoli, F., Dupraz, C., Gerin, M., Klein, U., Wieblinski, R. \& Brouillet, N. 1993, A\&A Suppl., 97, 887 
\bibitem[]{}Burton, M., Hollenbach, D. J., \& Tielens, A. G. G. M. 1990, ApJ,
365, 620
\bibitem[]{}Burton, M., Hollenbach, D. J., \& Tielens, A. G. G. M. 1992, ApJ,
399, 563
\bibitem[]{}B\"{u}ttgenbach, T.H., Keene, J., Phillips, T.G., \& Walker, C.K. 1992, ApJ,
        397, L15
\bibitem[]{}Carral, P., Hollenbach, D. J., Lord, S. D., Colgan, S. W. J., 
Haas, M. R., Rubin, R. H. \& Erickson, E. F. 1994, \apj, 423, 223
%\bibitem[]{}Colbert, J.W., et al. 1998, \apj, in press
\bibitem[]{}Crawford, M.K., Genzel, R., Townes, C.H., \& Watson, D.M. 1985, \apj, 291,
755
\bibitem[]{}Dalgarno, A. \& McCray, R. A. 1972, \araa, 10, 375
\bibitem[]{}Dixon, W. V. D., Hurwitz, M. \& Bowyer, S. 1998, \apj, 492, 569 
\bibitem[]{}Draine, B. T., \& Bertoldi, F. 1996, ApJ, 268, 269
\bibitem[]{}Draine, B. T., Roberge, W. G., \& Dalgarno, A. 1983, ApJ, 264, 485
\bibitem[]{}Draine, B. T., \& Sutin, B. 1987, ApJ, 320, 803 
\bibitem[]{}Ferland, G.J. 1996, University of Kentucky Department of Physics and
Astronomy Internal Report
\bibitem[]{}Fischer, J., et al. 1996, A\&A, 315, L97
\bibitem[]{}Fischer, J., et al. 1999, in preparation
\bibitem[]{}Freedman, W., et al. 1994, \apj, 427, 628
\bibitem[]{}Heiles, C. 1994, ApJ, 436, 720
\bibitem[]{}Helou, G. et al. 1996, BAAS, 189, 6704
\bibitem[]{}Hollenbach, D., \& McKee, C. F. 1979, ApJS, 41, 555
\bibitem[]{}Hollenbach, D., \& McKee, C. F. 1989, ApJ, 342, 306
\bibitem[]{}Hollenbach, D., Takahashi, T., \& Tielens, A. G. G. M. 1991, ApJ,
377, 192 (HTT) 
\bibitem[]{}Hollenbach, D., \& Tielens, A. G. G. M. 1997, \araa, 35, 179
\bibitem[]{}Hollenbach, D., \& Tielens, A. G. G. M. 1999, Rev.Mod.Phys., 71, 173
\bibitem[]{}Israel, F.P., Maloney, P.R., Geis, N., Herrmann, F., Madden, S.C., Poglitsch,
        A., \& Stacey, G.J. 1996, ApJ, 465, 738
\bibitem[]{}Jaquet, R. Staemmler, V.,  Smith, M. D., $\&$ Flower, D. R. 1992,
J. Phys.\ B, 25, 285
\bibitem[]{}Joy, M., Lester, D.F., \& Harvey, P.M. 1987, \apj, 319, 314
\bibitem[]{}Jura, M. 1974, \apj, 191, 375
\bibitem[]{}K\"oster, B., St\"orzer, H., Stutzki, J. \& Sternberg, A. 1994, \aa, 284, 545
\bibitem[]{}Le Bourlot, J. G., Pineau des For\^{e}ts, G., Roueff, E., Dalgarno, A., \& Gredel, R. 1993,
\apj, 449, 178
\bibitem[]{}Lequeux, J., Le Bourlot, J., Pineau des F\^{o}rets, G., Roueff, E., Boulanger,
        F., \& Rubio, M. 1994, A\&A, 292, 371
\bibitem[]{}Lord, S.D., Hollenbach, D. J., Haas, M. R., Rubin, R. H., Colgan, S. W. J. \& Erickson, E. F. 1996a, 
\apj, 465, 703
\bibitem[]{}Lord, S.D., et al. 1996b, A\&A, 315, L117
\bibitem[]{}Lord, S.D., et al. 1999, in preparation
\bibitem[]{}Luhman, M.L., Jaffe, D. T., Sternberg, A., Herrmann, F. \& Poglitsch, A. 1997a, \apj, 482, 298
\bibitem[]{}Luhman, M. L., Fischer, J. J., Satyapal. S., Wolfire, M. G., Stacey, G. J., Lord, S. D., Unger, S. J. \& Smith, H. A. 1997b, in Extragalactic Astronomy in the Infrared, eds. G. A. Mamon, T. X. Thuan \& J. T. Thanh Van (Paris: Editions Frontiers), 149
\bibitem[]{}Luhman, M.L., Satyapal, S., Fischer, J., Wolfire, M. G., Cox, P., Lord, S. D., Smith, H. A.,
Stacey, G. J., \& Unger, S. J. 1998, \apj, 504, L11
\bibitem[]{}Madden, S.C., Geis, N., Genzel, R., Herrmann, F., Jackson, J., Poglitsch, A.,
        Stacey, G.J., \& Townes, C.H. 1993, ApJ, 407, 579
\bibitem[]{}Madden, S.C., Poglitsch, A., Geis, N., Stacey, G.J., \& Townes, C.H. 1997, ApJ,
        483, 200
\bibitem[]{}Malhotra, S., Helou, G., Stacey, G., Hollenbach, D. J., Lord, S.,
Beichman, C. A.,  Dinerstein, H., Hunter, D. A., Lo, K. Y., Lu, N. Y., Rubin,
R. H., Silbermann, N., Thronson, H. A. \& Werner, M. W. 1997, ApJ, 491, 27
\bibitem[]{}Malhotra, S. et al. 1999, in preparation
\bibitem[]{}Maloney, P., \& Black, J. H. 1988, ApJ, 325, 389
\bibitem[]{}Maloney, P., \& Wolfire, M. 1997, in Proc. of the 170th IAU Symp.,
CO:         Twenty-Five Years of Millimeter-Wave Spectroscopy, eds. W.B.
Latter, S.J.E. Radford, P.R. Jewell, J.G. Mangum, \& J. Bally (Kluwer
Academic Pub.), 299 
\bibitem[]{}Mathis, J.S., Rumpl, W., \& Nordsieck, K.H. 1977, \apj, 217, 425
\bibitem[]{}McKee, C. F., \& Ostriker, J. P. 1977, \apj,218, 148
\bibitem[]{}McKee, C. F. 1989, \apj, 345, 782 
\bibitem[]{}Millar, T. J., Farquhar, P. R. A., \& Willacy, K. 1997,
A\&AS, 121, 139
\bibitem[]{}Moshir, M. et al. 1990, IRAS Faint Source Catalog, vol. 2
\bibitem[]{}Pak, S., Jaffe, D. T., Van Dishoeck, E. F., Johansson, L. E. B. \& Booth, R. S. 1998, \apj, 498, 735
\bibitem[]{}P\'equignot, D. 1990, A\&A, 231, 499
\bibitem[]{}Petuchowski, S. J. \& Bennett, C. L. 1993, \apj, 405, 591
\bibitem[]{}Poglitsch, A., Krabbe, A., Madden, S.C., Nikola, T., Geis, N.,
Johansson,         L.E.B., Stacey, G.J., \& Sternberg, A. 1995, ApJ, 454, 293
\bibitem[]{}Rothman, L.S. \ea 1987, Appl. Opt., 26, 4078 \bibitem[]{}Sakamoto,
S. 1996, ApJ, 462 215 
\bibitem[]{}Satyapal, S., Watson, D. M., Pipher, W. J., Forrest, J. L., Fischer, J., Greenhouse, M. A., Smith, H. A. \& Woodward, C. E. 1999, \apj, 516, 704
\bibitem[]{}Savage, B. D. \& Sembach, K. R. 1996, \araa, 34, 279
\bibitem[]{}Schr\"oder, K., Staemmler, V., Smith, M. D., Flower, D. R., \& Jaquet, 
R. 1991, J. Phys.\ B, 24, 2487
\bibitem[]{}Schilke, P., Carlstrom, J.E., Keene, J., \& Phillips, T.G. 1993, \apj,
417, L67
\bibitem[]{}Shull, J. M. \& Van Steenberg, M. E. 1985, \apj, 298, 268
\bibitem[]{}Spaans, M., Tielens, A. G. G. M., van Dishoeck, E. F., \& Bakes, E. L.
O. 1994, ApJ, 437, 270
\bibitem[]{}Stacey, G.J., Geis, N., Genzel, R., Lugten, J.B., Poglitsch, A., Sternberg, A.,
        \& Townes, C.H. 1991, ApJ, 373, 423
\bibitem[]{}Stark, A.A., Bolatto, A.D., Chamberlin, R.A., Lane, A.P., Bania, T.M., Jackson,
        J.M., \& Lo, K.-Y. 1997, ApJ, 480, L59
\bibitem[]{}Sternberg, A. \& Dalgarno, A. 1985, \apjs, 99, 565
\bibitem[]{}Sternberg, A. \& Dalgarno, A. 1989, \apj, 338, 197 
\bibitem[]{}Stutzki, J., et al. 1997, ApJ, 477, L33
\bibitem[]{}Telesco, C. M. \& Harper, D. A. 1980, \apj, 235, 392
\bibitem[]{}Tielens, A.G.G.M. \& Hollenbach, D.J. 1985a, ApJ, 291, 722 (TH85)
\bibitem[]{}Tielens, A.G.G.M. \& Hollenbach, D.J. 1985b, Icarus, 61, 40
\bibitem[]{}van Dishoeck, E.F., \& Black, J.H. 1986, \apjs, 62, 109  
\bibitem[]{}van Dishoeck, E.F., \& Black, J.H. 1988, in Molecular Clouds in the Milky Way
        and External Galaxies, eds. R.L. Dickman, R.L. Snell, \& J.S. Young
        (Springer-Verlag), 168
\bibitem[]{}van Dishoeck, E. F., \& Black, J. H. 1988, ApJ, 334, 771
\bibitem[]{}Viscuso, P. J. \& Chernoff, D. F. 1988, ApJ, 327, 364
\bibitem[]{}White, G.J., Ellison, B., Claude, S., Dent, W.R.F., Matheson, D.N. 1994, A\&A,
        284, L23
\bibitem[]{}Wild, W., Harris, A.I., Eckart, A., Genzel, R., Graf, U.U., Jackson, J.,
Russell, A.P.G., \& Stutzki, J. 1992, \aa, 265, 447
\bibitem[]{}Wolfire, M., Hollenbach, D. J., \& Tielens, A.G.G.M. 1989, ApJ, 344, 770
\bibitem[]{}Wolfire, M., Hollenbach, D. J., \& Tielens, A.G.G.M. 1993, ApJ, 402, 195
\bibitem[]{}Wolfire, M., Tielens, A.G.G.M., \& Hollenbach, D. 1990, ApJ, 358, 116 [WTH90]
\bibitem[]{}Wolfire, M. G., Hollenbach, D. McKee, C. F., Tielens, A. G. G. M.,
\& Bakes, E. L. O. 1995, ApJ, 443, 152
\bibitem[]{}Wright, E. L., et al. 1991, \apj, 381, 200
\bibitem[]{}Young, J.S., Kenney, J.D., Tacconi, L., Claussen, M.J., Huang,Y.-L., Tacconi-Garman, L., Xie, S. \& Schloerb, F.P. 1986, \apj, 311, L17
\bibitem[]{}Young, J.S., Allen, L., Kenney, J.D.P., Lesser, A. \& Rownd, B. 1996, \aj, 112, 1903
\end{thebibliography}
\end{document}